\documentclass[acmsmall,manuscript]{acmart}
\usepackage{booktabs}
\usepackage{graphicx}
\usepackage{epstopdf}
\usepackage{colortbl} 
\usepackage{xcolor}
\usepackage{array}
\usepackage{enumerate}
\usepackage{xspace}

\usepackage{soul}

\usepackage{amssymb}
\usepackage{amstext}
\usepackage{amsmath}
\usepackage{amsthm}
\usepackage{algorithm}
\usepackage{algorithmic}
\usepackage{subfigure}
\usepackage{multirow}
\usepackage{makecell}
\usepackage{comment}

\AtBeginDocument{
  }
\setcopyright{acmcopyright}
\copyrightyear{2018}
\acmYear{2018}
\acmDOI{XXXXXXX.XXXXXXX}

\acmJournal{JACM}
\acmVolume{37}
\acmNumber{4}
\acmArticle{111}
\acmMonth{8}

\newcommand{\ours}[0]{GraphTransfer\xspace}

\begin{document}
\title{\ours: A Generic Feature Fusion Framework for Collaborative Filtering}

\author{Jiafeng Xia}
\authornote{Also with Shanghai Key Laboratory of Data Science, Fudan University, Shanghai, China.}
\email{jfxia19@fudan.edu.cn}
\affiliation{
  \institution{School of Computer Science, Fudan University}
  \city{Shanghai}
  \country{China}
}

\author{Dongsheng Li}
\authornote{Also with School of Computer Science, Fudan University, Shanghai, China.}
\email{dongshengli@fudan.edu.cn}
\affiliation{
  \institution{Microsoft Research Asia}
  \city{Shanghai}
  \country{China}
}

\author{Hansu Gu}
\authornote{Corresponding author.}
\email{hansug@acm.org}
\affiliation{
  \institution{Amazon.com}
  \city{Seattle}
  \country{USA}
}

\author{Tun Lu}
\authornotemark[1]
\authornotemark[3]
\email{lutun@fudan.edu.cn}
\affiliation{
  \institution{School of Computer Science, Fudan University}
  \city{Shanghai}
  \country{China}
}

\author{Ning Gu}
\authornotemark[1]
\email{ninggu@fudan.edu.cn}
\affiliation{
  \institution{School of Computer Science, Fudan University}
  \city{Shanghai}
  \country{China}
}

\renewcommand{\shortauthors}{Jiafeng Xia et al.}

\begin{abstract}
  Graph Neural Networks (GNNs) have demonstrated effectiveness in collaborative filtering tasks due to their ability to extract powerful structural features. However, combining the graph features extracted from user-item interactions and auxiliary features extracted from user genres and item properties remains a challenge. Currently available fusion methods face two major issues: 1) simple methods such as concatenation and summation are generic, but not accurate in capturing feature relationships; 2) task-specific methods like attention mechanisms and meta paths may not be suitable for general feature fusion. To address these challenges, we present \ours, a simple but universal feature fusion framework for GNN-based collaborative filtering. Our method accurately fuses different types of features by first extracting graph features from the user-item interaction graph and auxiliary features from users and items using GCN. The proposed cross fusion module then effectively bridges the semantic gaps between the interaction scores of different features. Theoretical analysis and experiments on public datasets show that \ours outperforms other feature fusion methods in CF tasks. Additionally, we demonstrate the universality of our framework via empirical studies in three other scenarios, showing that \ours leads to significant improvements in the performance of CF algorithms.
\end{abstract}



\maketitle

\section{Introduction}
In the era of information overload, recommender systems play a pivotal role  since it helps people reduce the time wasted in finding useful information by recommending relevant items to users~\cite{rs1, rs2, rs3, rs4, rs5, fire, higsp, fagsp}. In recent years, graph neural network based recommendation algorithms have achieved state-of-the-art performance in collaborative filtering tasks, including sequential recommendation~\cite{jodie,deepcoevolve,fire,igcn,seqrecref1, neufilter}, social recommendation~\cite{social1,social2}, POI recommendation~\cite{poi1,poi2, poi3}, etc., due to their powerful graph feature (including user preference and structural feature) extraction ability over user-item interaction graph~\cite{gcn,gin,powerfulsgnn}.  Inspired by CNN~\cite{krizhevsky2012imagenet}, GCN~\cite{gcn} extracts the local structure information of nodes through the convolutions defined on the graph, which has achieved the most promising results among GNN-based CF methods~\cite{lightgcn,impgcn, lrgccf}. Specifically, each convolution of GCN contains an aggregation operation and an update operation. Through these two operations, node can obtain the features of neighboring nodes based on the connectivity between nodes. With continuously aggregating and updating the representations of nodes on the user-item interaction graph, it can capture more graph feature from farther neighboring nodes to accurately model the features of users/items and improve the accuracy of recommendations~\cite{lightgcn,impgcn,lrgccf,gfcf,ghcf,li2020learning, igcn}.

In the realm of recommendation algorithm research, utilizing multiple types of features for user and item preference modeling has been shown to improve accuracy by providing a multi-dimensional understanding~\cite{roy2021multi}. As diverse features convey orthogonal and supplementary information, this approach enables the construction of more complete user profiles and item characteristics, leading to enhanced algorithm performance. 
In addition to graph feature in user-item interactions, another type of feature used for describing users and items is auxiliary feature, which includes characteristics such as user genres and item properties. However, it is a challenging problem to fuse the graph feature and the auxiliary feature to improve the performance of CF algorithms. Graph feature and auxiliary feature are derived from separate spaces and convey distinct meanings, with the former describing users and items from a preference perspective and the latter from an inherent perspective. The fusion process therefore presents a significant challenge in preserving the original information from both features. Existing methods adopted concatenation~\cite{graphrec}, summation~\cite{summationfusion, cpf}, attention mechanism~\cite{attention} or meta paths~\cite{hin,han} to fuse different types of features. 
Concatenation and summation based methods, which merely concatenate and add different features in the fusion process, are too simple to fully capture the non-linearity between different types of features, and thus cannot optimally fuse them. Although attention based and meta path based methods provide an improvement in capturing non-linearity between features compared to concatenation and summation based methods, researchers usually make some modifications and improvements on the attention mechanism in order to better adapt them to specific tasks~\cite{niu2021review}, and the design of meta paths relies heavily on specialized domain knowledge to manually design task-specific feature extraction patterns, making them task-specific and cannot fuse different types of feature in a universal manner. 
In summary, these methods are faced with two main challenges: (1) they fail to accurately and effectively capture the relations of different types of features to fully fuse them; (2) they are task-specific and cannot generally fuse different features, which will limit the potential for the performance improvement of recommendation algorithms.

To this end, we propose \ours, a simple but generic feature fusion framework for GNN-based CF methods, which can fully fuse different types of features using a universal manner. Firstly, with the help of existing graph embedding methods as the backbone module, we extract the graph feature of users and items from user-item interaction graph. Then, we design a graph neural network based auxiliary feature extraction module that can effectively learn auxiliary feature of users and items from their auxiliary information based on the user-user interaction graph and item-item interaction graph constructed using user-item interaction graph. Finally, \ours realizes the fusion of different types of features through the carefully designed cross fusion module in a manner of ``learn to fuse'', which minimizes the gap between the interaction scores produced by a novel cross-dot-product mechanism of different features to bridge the semantic gaps in a universal manner. Theoretical analysis shows the effectiveness of cross fusion module that it considers the both information of graph feature and auxiliary feature when updating representations of users and items compared with other fusion methods. Extensive experiments on three public datasets show that \ours can achieve better performance compared with the state-of-the-art methods. Meanwhile, \ours can also alleviate the over-smoothing issue~\cite{oono2019graph} of GCN based CF methods due to the introduction of auxiliary feature. 
Moreover, we analyze the universality of the proposed framework \ours by fusing different types of features in three other scenarios and the experimental results show that \ours can effectively fuse different types of features and improve the performance of CF algorithms with a considerable margin.

The main contributions of this work are summarized as follows:
\begin{itemize}
    \item We design an auxiliary feature extraction module based on graph convolutional network to efficiently extract auxiliary features of users and items from the user-user and item-item interaction graphs constructed by user-item interaction graph;
    \item We propose a generic feature fusion framework, \ours, to effectively fuse auxiliary feature and graph feature in a universal manner through a carefully designed cross fusion module, and the theoretical analysis proves the effectiveness of the cross fusion module compared with other fusion methods;
    \item We conduct extensive experiments and the results show that \ours can achieve better performance compared with the state-of-the-art methods and alleviate the over-smoothing issue of GCN-based CF methods by introducing auxiliary feature;
    \item We further conduct a universality experiment of \ours, and the results show that it has high universality and is able to effectively fuse structural and text features from user reviews, as well as features from different time periods in sequential recommendation models and features extracted by different models.
\end{itemize}

\section{Related Work}
\subsection{GNN based Recommendation}
In collaborative filtering, graph neural network based recommendation algorithms have recently attracted extensive attentions from  researchers~\cite{survey1,survey2,gnncftref1, gnncfref2}. They have  achieved excellent performance in various scenarios, such as sequential recommendation~\cite{deepcoevolve,jodie,stpudgat,liu2020inter,surge, retagnn} and social recommendation~\cite{social1,social2,wu2019neural,social3}. In addition, they can also make explanations for the recommendation results, so as to improve the interpretability of the recommendations~\cite{interpret1,interpret2,interpret3,interpret4,zhang2020explainable,fu2020fairness, lyu2022knowledge}. Graph neural networks can extract and model the structural feature of users and items from the user-item interaction graph through the process of constantly aggregating and updating their representations, thus improving the accuracy of recommendation results. GCMC~\cite{gcmc} designed an Encoder-Decoder architecture on the user-item interaction graph to predict possible ratings. NGCF~\cite{ngcf} explicitly encoded the collaborative signal in the form of high-order connectivities by performing feature propagation over the user-item interaction graph, thus achieving promising performance. LightGCN~\cite{lightgcn} improved the accuracy of the recommendation results by removing two useless operations in NGCF, which are feature transformation and nonlinear activation. IMP-GCN~\cite{impgcn} extracted features from subgraphs which consist of users with similar interests and their interacted items to improve the  performance. GraphRec~\cite{graphrec} took both the users’ social relationship and the user's interactions into account to improve the performance.

However, due to the over-smoothing~\cite{oono2019graph} issue of graph neural networks, the number of layers in GNN is greatly limited, which may restrict the performance of GNN-based recommendation algorithm. A lot of works tried to alleviate the over-smoothing issue of graph neural networks~\cite{liu2020towards,bo2021beyond,lrgccf, grato, svdgcn}. The LR-GCCF~\cite{lrgccf} method removes non-linearities and introduces a residual network to ease model training and alleviate over-smoothing issues in graph convolution aggregation operations. Additionally, it is a linear model, making it easy to train, and has been proven to achieve better efficiency and effectiveness. Different from the above works, we alleviate the over-smoothing issue by introducing auxiliary information using the proposed \ours framework.

\subsection{Feature Fusion in Collaborative Filtering}

By utilizing multiple types of features, the accuracy of recommendations can be significantly improved in comparison to using a single type of feature~\cite{roy2021multi}, since different features depict the characteristics of user and item from different perspectives, they are orthogonal and complementary, integrating them to jointly model the representations of users and items can make the modeling more precise, thus achieving better performance. Fusing multiple types of features to improve the performance of collaborative filtering algorithms is a challenging task as different features originate from various spaces and possess distinct semantics. It can often present difficulties in retaining all of the original information from those features during the fusion process. Existing methods achieve the fusion of feature through concatenation~\cite{graphrec}, summation~\cite{summationfusion,cpf}, attention mechanism~\cite{amgcn,attention} or meta paths~\cite{hin_survey,han}. GraphRec~\cite{graphrec} proposed to concatenate two different features: user latent representations from item space and social space, respectively, to precisely model user characteristics. Yang et al.~\cite{cpf} proposed to sum different features using learnable weights to achieve feature fusion. Wang et al.~\cite{amgcn} achieved the fusion of feature through an attention mechanism among three different kinds of features extracted by GNNs. The HAN method~\cite{han} realized the fusion by devising multiple meta paths to learn the features of user and item, and more sophisticated meta paths lead to better performance. 

Note that concatenation or summation based methods, which merely concatenate or add different features, are too simple to fully capture the non-linearity between different features, and thus they cannot optimally fuse different types of features. While attention or meta path based methods can capture the non-linearity between different features, they are task-specific and cannot fuse different types of feature in a universal manner. Though the architecture of attention mechanisms is generally the combination of values and weights calculated from the matching degree between query and keys, researchers often make modifications and improvements to better suit specific tasks~\cite{niu2021review, ffanet, saffnnet, attention, amgcn, affniaff}, e.g., AMVF~\cite{attention} adopts separate learnable weights which sum to 1 as attention weights to fuse features for cross-domain recommendation, AM-GCN~\cite{amgcn} adopts shared weights as key to calculate attention weights and combine different types of features for node classification. AFF and iAFF~\cite{affniaff} design a more sophisticated structure MS-CAM, which consists of some convolution operations, to fuse features of different scales for image classification and semantic segmentation. For meta path based methods, they rely heavily on specialized domain knowledge to manually design sophisticated features, and they need to design different types of meta paths for different tasks to improve the accuracy of feature extraction. Therefore, these methods are faced with two main challenges: (1) they cannot accurately capture the relationships between different types of features and fully fuse them, (2) they are specific to certain tasks, making them unsuitable for general feature fusion. These challenges limit the potential for improvement in the performance of recommendation algorithms.

\section{Preliminaries}

\subsection{Notations} 
Let $\mathcal{G}$ be a graph with a node set $\mathcal{V}$ and an edge set $\mathcal{E}$, and $|\mathcal{V}|=l$. Mathematically, the graph $\mathcal{G}$ can be represented as an adjacency matrix $\mathbf{A}\in\mathbb{R}^{l\times l}$, and if there is an edge between node $i$ and node $j$, then $\mathbf{A}_{ij}=1$, otherwise, $\mathbf{A}_{ij}=0$. In recommender systems, $\mathcal{G}$ is usually a bipartite graph, where there are two types of nodes, i.e., users and items. We denote the set of user nodes as $\mathcal{U}$, and the set of item nodes as $\mathcal{I}$, $|\mathcal{U}|=n$ and $|\mathcal{I}|=m$. Similar to the adjacency matrix $\mathbf{A}$, we use a user-item interaction matrix $\mathbf{R}\in\mathbb{R}^{n\times m}$, which is generated from user interactions $\mathcal{O}=\{(u,i,r)|u\in\mathcal{U}, i\in \mathcal{I}\}$, to describe the relationship between users and items, if there is an interaction $r$ (either a binary value or a numerical value) between user $u$ and item $i$, then $\mathbf{R}_{ui}=r$, otherwise $\mathbf{R}_{ui}=0$. In addition to interactive features present in user interactions, both users and items possess their own, unique features, referred to as auxiliary features, e.g. user's age and occupation and item's category and popularity. For each user, suppose there are $p$ auxiliary features $\{f_1, f_2,\cdots,f_p\}$, and if the value of some feature $f_i$ is missing, we pad it with a blank token. Then we transform each auxiliary feature to an one-hot encoding, and concatenate these $p$ one-hot encodings to the primary user auxiliary feature $\mathbf{X}\in\mathbb{R}^{n\times d_1}$, where $d_1$ is the dimension of primary user auxiliary feature. Similarly, for each item, we use blank tokens to pad the missing values in the features, and generate the primary item auxiliary feature $\mathbf{Y}\in\mathbb{R}^{m\times d_2}$, where $d_2$ is the dimension of primary item auxiliary feature. Table~\ref{tab:notation} shows the notations that are used throughout the paper.

\begin{table*}[h!]
\renewcommand{\arraystretch}{1.3}
\caption{The notations used in this manuscript.}
\label{tab:notation}
\centering
\small
\begin{tabular}{cp{10cm}}
\toprule
{\bf Notation} & {\bf Description} \\
\midrule
$\mathcal{G}, \mathcal{V}, \mathcal{E}$ & The graph $\mathcal{G}$ and its corresponding node set $\mathcal{V}$ and edge set $\mathcal{E}$\\
$\mathbf{A}$&The adjacency matrix of graph $\mathcal{G}$\\
$\mathcal{U}, \mathcal{I}$&The user set and the item set.\\
$\mathcal{N}(i)$ & The 1-hop neighboring node set of node $i$\\
$\mathcal{S}$ & The set of sub-graphs that consists of users with similar interests and their interacted items\\
$\mathcal{N}^s(i)$ & The 1-hop neighboring node set of node $i$ in a sub-graph $s\in\mathcal{S}$\\
$d_i$&The degree of node $i$\\
$\mathbf{R}$&The user-item interaction matrix\\
$\mathcal{O}$&The user interactions that contains multiple triplets $(u,i,r)$ corresponding to users' interaction records\\
$\mathbf{s}_i$ &The structural feature of node $i$\\
$\mathbf{s}_{i}^{(k)}$ &The structural feature of node $i$ at $k$-th layer\\
$\mathbf{s}_{\mathcal{N}(i)}^{(k)}$ &The aggregated structural feature of neighborhood of node $i$\\
\midrule
$\mathbf{X}, \mathbf{Y}$&The primary auxiliary feature of user and item.\\
$\mathbf{U}$&The user-user interaction matrix\\
$\mathbf{g}_u, \mathbf{g}_i$ &The graph feature of user $u$, of item $i$\\
$\mathbf{g}_{u}^{(k)}, \mathbf{g}_{i}^{(k)}$ &The graph feature of of user $u$ and item $i$ at $k$-th layer\\
$\mathbf{a}_u, \mathbf{a}_i$&The auxiliary feature of user $u$ and item $i$\\
$r^g_{u,i}$&The interaction score between graph feature of user $u$ and item $i$\\
$r^a_{u,i}$&The interaction score between auxiliary feature of user $u$ and item $i$\\
$r^{c1}_{u,i}, r^{c2}_{u,i}$&The interaction scores between graph feature of user $u$ and of auxiliary feature item $i$, and auxiliary feature of user $u$ and of graph feature item $i$ respectively\\
$\mathbf{h}_u^{t}, \mathbf{h}_i^{t}$&The temporal feature of user $u$ and item $i$ at time period $t$\\
\bottomrule
\end{tabular}
\end{table*}

\subsection{Graph Neural Networks}

\label{sec:gnn}
Graph neural network has powerful structural feature extraction ability on graph data.  Given a graph $\mathcal{G}=(\mathcal{V}, \mathcal{E}) $, graph neural network models the structural feature of nodes by performing the following two operations at each layer:
\begin{align}
    \mathbf{s}^{(k)}_{\mathcal{N}(i)}&=\mathrm{AGG}(\{\mathbf{s}^{(k-1)}_j,~\forall~j\in\mathcal{N}(i)\}),&\textbf{(Aggregation)}\\
    \mathbf{s}^{(k)}_i&=\mathrm{UPD}(\{\mathbf{W}^{(k)}_1\mathbf{s}^{(k)}_{\mathcal{N}(i)}, \mathbf{W}^{(k)}_2\mathbf{s}^{(k-1)}_i\}),&\textbf{(Update)}
\end{align}
where $\mathbf{s}^{(k-1)}_i$ and $\mathbf{s}^{(k)}_i$ are the representations containing structural feature of node $i$ at the $(k-1)$-th layer and $k$-th layer, $\mathbf{s}^{(k)}_{\mathcal{N}(i)}$ is the aggregated representation of neighborhood of node $i$ at layer $k$ and $\mathcal{N}(i)$ is the 1-hop neighboring node set of node $i$, $\mathbf{W}^{(k)}_1$ and $\mathbf{W}^{(k)}_2$ are learnable parameters at layer $k$, and we neglect bias term for simplicity, $\mathrm{AGG}(\cdot)$ and $\mathrm{UPD}(\cdot)$ are aggregation function and update function, which can be a summation, average, multi-layer perceptron (MLP) or other operations. The process of updating the representation of each target node involves aggregating the features of neighboring nodes and its own features, as determined by the connectivity between the nodes, which enables the node to capture the structural features of the graph.

However, a single aggregation operation and update operation can only get the structural feature of 1-hop neighbors for each node. Graph neural network obtains the structural feature of farther neighbors (e.g., 2-hop or 3-hop neighbors) by stacking multiple layers, each of which contains an aggregation operation and an update operation, so as to model the structural feature of node more accurately. Next, we introduce four state-of-the-art graph neural networks that can extract structural feature on graph data.

\subsubsection{GIN}
GIN~\cite{gin}  is a highly effective approach for extracting structural features by utilizing a summation operation in the aggregation step and a MLP in the update step, which shows promising performance of discriminating isomorphic graphs in graph learning task. When applied to recommendation task, GIN updates user representation as follows:
\begin{equation}
\mathbf{s}_u^{(k)}=\mathrm{MLP}^{(k)}\left((1+\epsilon^{(k)})\cdot \mathbf{s}_u^{(k-1)}+\sum_{i\in\mathcal{N}(u)}\mathbf{s}_i^{(k-1)}\right),
\end{equation}
where $\mathrm{MLP}^{(k)}$ is the MLP in the  $k$-th layer, and $\epsilon^{(k)}$ is a learnable parameter or a fixed scalar that can be used to discriminate the feature of target node and the aggregated features from the neighborhood. Item representations can be
obtained in the same manner.

\subsubsection{LR-GCCF}
LR-GCCF~\cite{lrgccf} is a linear model that can handle large datasets with high efficiency and effectiveness. The user representation can be obtained as follows:
\begin{equation}
    \mathbf{s}_u^{(k)}=\mathbf{W}^{(k-1)}\left(\frac{1}{d_u}\mathbf{s}_u^{(k-1)}+\sum_{i\in\mathcal{N}(u)}\frac{1}{d_i\times d_u}\mathbf{s}_i^{(k-1)}\right),
\end{equation}
where $d_u$ and $d_i$ are the degrees of user $u$ and item $i$ respectively. Item representations can be obtained in a similar manner.

\subsubsection{LightGCN}
LightGCN~\cite{lightgcn} achieves high efficiency by removing two useless operations, i.e., feature transformation and nonlinear activation, from traditional graph neural networks~\cite{ngcf}. The user representation can be obtained as follows:
\begin{equation}
\begin{aligned}
\mathbf{s}_u^{(k)}&=\sum_{i\in\mathcal{N}(u)}\frac{1}{\sqrt{|\mathcal{N}(u)|}\sqrt{|\mathcal{N}(i)|}}\mathbf{s}_i^{(k-1)},\\
    \mathbf{s}_u &= \sum_{k=0}^K\alpha_k\mathbf{s}_u^{(k)},
\end{aligned}
\end{equation}
where $\alpha_k$ denotes the importance of the $k$-th layer representation, and it can be a hyper-parameters or a learnable parameter. $\mathbf{s}_u$ is the final representation of user $u$ that aggregates information from all $K$ layers. Item representations can be obtained similarly.

\subsubsection{IMP-GCN}
IMP-GCN~\cite{impgcn} can model high-order structural features of users and items by introducing user sub-graphs. The item representation can be obtained in the following manner:
\begin{equation}
\begin{aligned}
\mathbf{s}_{i}^{(k)}&=\sum_{s\in\mathcal{S}}\sum_{u\in\mathcal{N}^s(i)}\frac{1}{\sqrt{|\mathcal{N}^s(i)|}\sqrt{|\mathcal{N}^s(u)|}}\mathbf{s}_u^{(k-1)},\\
\mathbf{s}_i&= \sum_{k=0}^K\alpha_k\mathbf{s}_i^{(k)},
\end{aligned}
\end{equation}
where $\mathbf{s}_{i}^{(k)}$ is the representation of item $i$ at the $k$-th layer. $\mathcal{S}$ is the set of sub-graphs and $\mathcal{N}^s(u)$ and $\mathcal{N}^s(i)$ are the 1-hop neighboring nodes of user $u$ and item $i$ in sub-graph $s\in\mathcal{S}$. $\mathbf{s}_i$ is the final representation of item $i$ that aggregates information from all $K$ layers. The calculation of user representation is the same as that in LightGCN. 

\section{The \ours Framework}
\subsection{Overview}
\begin{figure*}[!t]
\centering
\includegraphics[width=1.0\textwidth]{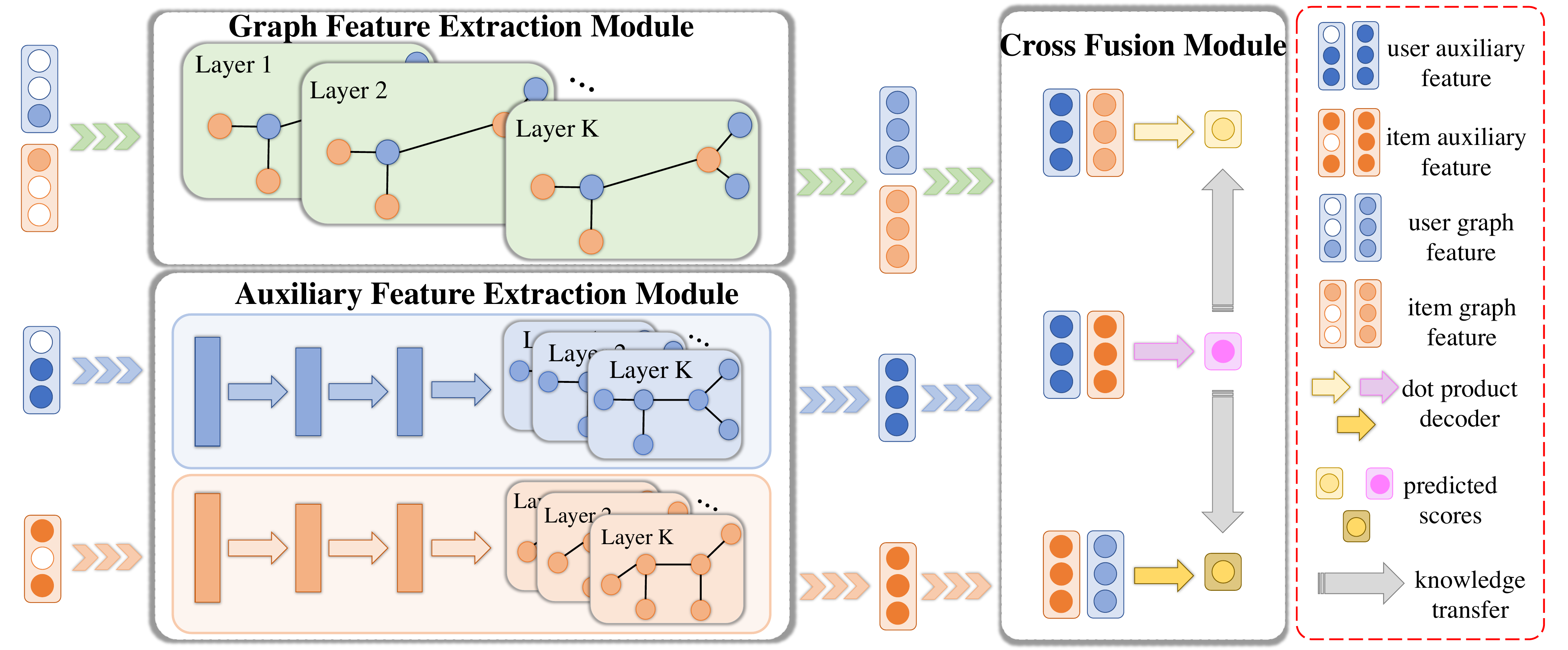}
\caption{The architecture of the proposed \ours framework, which consists of three modules: Graph Feature Extraction Module, Auxiliary Feature Extraction Module and Cross Fusion Module.}
\label{fig:model_stru}
\end{figure*}

Figure~\ref{fig:model_stru} shows the architecture of the proposed \ours framework, which consists of the following three key components: 
\begin{enumerate}[1.]
    \item[$\bullet$] {\bf Graph feature extraction module}, which can extract the graph features (including user preference and structural feature) of users and items through the graph embedding methods from the user-item interaction graph.
    \item[$\bullet$] {\bf Auxiliary feature extraction module}, which can extract the auxiliary features of users and items from their auxiliary information through graph convolutional network based on the user-user and item-item interaction graphs generated from user-item interaction graph.
    \item[$\bullet$] {\bf Cross fusion module}, which can fuse graph and auxiliary features of users and items through a ``learn to fuse'' approach, which minimizes the gap between the interaction scores produced by a novel cross-dot-product mechanism of different features to bridge the semantic gaps in a universal manner.
\end{enumerate}

\subsection{Graph Feature Extraction Module}
In this paper, graph feature represents the user preferences from user-item interactions and the structural feature from the interaction graph, which can be learned by existing GNN methods as mentioned in Section~\ref{sec:gnn}. Thus, we use the graph embedding modules in existing GNN-based CF methods as the graph feature extraction module. Take LightGCN~\cite{lightgcn} as an example, we can obtain the representations of user $u$ ($\mathbf{g}_u$) and item $i$ ($\mathbf{g}_i$) that contain graph feature as follows:
\begin{equation}
\begin{aligned}
    \mathbf{g}_u= \sum_{k=0}^K\alpha_k\mathbf{g}_u^{(k)},~\mathbf{g}_u^{(k)}=\sum_{j\in\mathcal{N}(u)}\frac{1}{\sqrt{|\mathcal{N}(u)|}\sqrt{|\mathcal{N}(j)|}}\mathbf{g}_{j}^{(k-1)},\\
    \mathbf{g}_i= \sum_{k=0}^K\alpha_k\mathbf{g}_i^{(k)},~\mathbf{g}_i^{(k)}=\sum_{v\in\mathcal{N}(i)}\frac{1}{\sqrt{|\mathcal{N}(v)|}\sqrt{|\mathcal{N}(i)|}}\mathbf{g}_{v}^{(k-1)},
\end{aligned}
\end{equation}
where $\mathbf{g}_u^{(k)}$ and $\mathbf{g}_i^{(k)}$ are the representations of user $u$ and item $i$ at the $k$-th layer respectively. We follow LightGCN~\cite{lightgcn} to use the BPR loss~\cite{bpr} to optimize the model parameters in this module:
\begin{equation}
    \mathcal{L}_{g}=-\sum_{(u,i^+)\in\mathcal{O}, i^-\notin\mathcal{N}(u), i^-\in\mathcal{I}}{\ln \sigma(r^g_{u,i^+}-r^g_{u,i^-})}+\lambda||\mathbf{G}^{(0)}||^2_F, 
\end{equation}
where $i^+$ is a positive item that user $u$ interacted with, and $i^-$ is a randomly sampled negative item that user $u$ has never interacted with. $r^g_{u,i^+}$ represents the interaction score between user $u$ and positive item $i^+$: $r^g_{u,i^+}={\mathbf{g}_u}^T\mathbf{g}_{i^+}$, similarly, $r^g_{u,i^-}$ represents the interaction score between user $u$ and negative item $i^-$: $r^g_{u,i^-}={\mathbf{g}_u}^T\mathbf{g}_{i^-}$. $\sigma(\cdot)$ is an activation function. $\mathbf{G}^{(0)}$ is the embedding of users and items in the 0-th layer, and $\lambda$ is the regularization coefficient.

\subsection{Auxiliary Feature Extraction Module}
Auxiliary features, such as user genres (e.g., age, occupation) and item properties (e.g., category, popularity), supplement traditional collaborative filtering by providing different perspectives on user and item characteristics. Unlike graph features that focus on preference, auxiliary features assist in modeling the fundamental characteristics of users and items. To this end, we propose an auxiliary feature extraction module to extract the auxiliary features to model the characteristics of users and items, as shown in the bottom left part of Figure~\ref{fig:model_stru}. The process of extracting auxiliary features for both users and items is similar, so for the sake of clarity, we will use the example of extracting user auxiliary features to show how the auxiliary feature extraction module operates.

A user's ($u$) primary auxiliary feature, represented as $\mathbf{x}_u$, may contain a large number of dimensions after being concatenated from $p$ one-hot encodings. To mitigate this, we use a MLP to reduce the dimension 
as follows:
\begin{equation}
    \mathbf{a}'_u=\mathbf{W}^{(L)}(\cdots(\sigma(\mathrm{BN}(\mathbf{W}^{(2)}(\sigma(\mathrm{BN}(\mathbf{W}^{(1)}\mathbf{x}_u+\mathbf{b}^{(1)})))+\mathbf{b}^{(2)}))))+\mathbf{b}^{(L)},
\label{eq:aux_encode}
\end{equation}
where $L$ is the number of layers in the MLP, $\{\mathbf{W}^{(1)}, \mathbf{b}^{(1)}, \mathbf{W}^{(2)}, \mathbf{b}^{(2)}, \cdots, \mathbf{W}^{(L)}, \mathbf{b}^{(L)}\}$ are learnable parameters of weights and biases of all $L$ layers, $\mathrm{BN}(\cdot)$~\cite{batchnorm} is the batch normalization which is used to stabilize the model training. $\sigma(\cdot)$ denotes the activation function, and we use ReLU in this paper. Thus, we obtain a compact and informative $d$-dimensional auxiliary feature of user $u$ ($\mathbf{a}'_u$).

To improve the representation of user's auxiliary feature, we adopt a Graph Convolutional Network (GCN) that retrieves information from users related to user $u$. This enhances the expressivity of user's auxiliary feature by aggregating more relevant data. Instead of using the user-item interaction graph $\mathbf{R}$, we employ GCN over the user-user interaction graph $\mathbf{U}$, which is generated from the user-item interaction graph, to aggregate information. Compared with $\mathbf{R}$, using $\mathbf{U}$ is more effective to improve the modeling of user auxiliary feature because it contains more direct information about the relationship among users. The user-user interaction graph $\mathbf{U}$ is constructed through the user-item interaction matrix $\mathbf{R}$ as follows:
\begin{equation}
    \mathbf{U}_{uv}=\frac{\mathbf{R}_u^T\mathbf{R}_v}{||\mathbf{R}_u||_2^2||\mathbf{R}_v||_2^2}~~~(u=1,2,\cdots,n; v=1,2,\cdots,n),
\label{eq:aux_graph1}
\end{equation}
where $\mathbf{R}_u$ is the $u$-th row of $\mathbf{R}$ and $||\cdot||_2$ is the $L_2$-norm. We use cosine function to measure the similarity between two users, however, other methods such as Jaccard Similarity can also be used to measure the similarity between two users. To reduce computation complexity and improve efficiency, we apply a threshold $\epsilon$ to make matrix $\mathbf{U}$ sparse. This also helps in removing irrelevant users that may negatively impact user feature modeling. Specifically, for each user, we retain only a small number of highly similar users and ignore those with low similarity, as determined by comparison of the similarity between users and the threshold $\epsilon$ as follows:
\begin{equation}
    \mathbf{U}_{uv}=
    \begin{cases}
        \mathbf{U}_{uv} &, \text{if} ~~\mathbf{U}_{uv}\ge\epsilon,\\
        0 &, \text{otherwise.}
\end{cases}
\label{eq:aux_graph2}
\end{equation}
Then, to improve representation learning of user's auxiliary feature, we apply a $K$-layer GCN on the sparse user-user interaction graph $\mathbf{U}$ as follows:
\begin{align}
\mathbf{a}^{(k)}_u=f\left(\sum_{v=1}^n \mathbf{U}_{uv}\mathbf{a}^{(k-1)}_v\right),
\label{eq:aux_gcn}
\end{align}
where $\mathbf{a}^{(k)}_u$ and $\mathbf{a}^{(k-1)}_u$ are the representations of user $u$ at the $k$-th and $(k-1)$-th layer, and we have $\mathbf{a}^{(0)}_u=\mathbf{a}'_u$. $f(\cdot)$ is a MLP with a batch normalization and an activation function in each layer. Note that we consider the self-loop of user $u$ when calculating $\mathbf{a}^{(k)}_u$, since $\mathbf{U}_{uu}=1$. The final auxiliary feature of user $u$ is $\mathbf{a}_u=\mathbf{a}^{(K)}_u$. Similarly, the final auxiliary feature of item $i$ is $\mathbf{a}_i = \mathbf{a}^{(K)}_i$.

\subsection{Cross Fusion Module}

Different features provide unique information from diverse sources or perspectives that, when combined, creates a comprehensive depiction of users and items. The integration of these features leads to a more accurate modeling process, resulting in improved recommendation performance. 
However, existing feature fusion methods face with two main challenges: 1) they fail to accurately capture the relationships between different feature types, hindering the fusion process and 2) they are task-specific and cannot be generalized to different applications. Thus, we propose the cross fusion module, a generic feature fusion module to fuse different types of features, as shown in the right part of Figure~\ref{fig:model_stru}.

Given the graph features of user $u$ and item $i$ ($\mathbf{g}_u$ and $\mathbf{g}_i$) extracted by graph feature extraction module and their corresponding auxiliary features ($\mathbf{a}_u$ and $\mathbf{a}_i$) extracted by auxiliary feature extraction module, we first calculate three types of interaction scores $r^a_{ui}$, $r^{c1}_{ui}$ and $r^{c2}_{ui}$ using a novel cross-dot-product as follows:
\begin{equation}
    r^a_{ui} = {\mathbf{a}_u}^T\mathbf{a}_i,~~
    r^{c1}_{ui} = {\mathbf{g}_u}^T\mathbf{a}_i,~~
    r^{c2}_{ui} = {\mathbf{a}_u}^T\mathbf{g}_i.
\end{equation}
Note that $r^a_{ui}$ is calculated between the graph features of user $u$ and item $i$, while both $r^{c1}_{ui}$ and $r^{c2}_{ui}$ are calculated between the features of user $u$ and item $i$ from different sources. We use $r^{c1}_{ui}$ and $r^{c2}_{ui}$ to capture the relationship between graph feature and auxiliary feature. These two metrics allow us to effectively fuse these two types of features, which can be helpful in improving the performance of the recommendation algorithms. Other interaction functions can also be applied to obtain the interaction scores between different features in a cross decoding manner. However, we choose to use the dot product method as it offers a high level of computational efficiency since no additional learnable parameters are introduced. 

We propose a new approach to feature fusion that differs from traditional methods which either attempt to minimize the gap between features from different sources~\cite{spacefusion1,spacefusion2} or map them into a shared feature space~\cite{summationfusion}. Instead, our method minimizes the gap between interaction scores. In this design, we focus on the consistency of prediction results rather than the consistency of feature space, which may achieve more effective feature fusion because our strategy is better in line with the final goal --- high recommendation accuracy. More specifically, we fuse the graph feature and the auxiliary feature by minimizing the following losses:
\begin{equation}  
\mathcal{L}_{c1}=\sum_{(u,i)\in\mathcal{O}}(r^a_{ui}-r^{c1}_{ui})^2,\quad
\mathcal{L}_{c2}=\sum_{(u,i)\in\mathcal{O}}(r^a_{ui}-r^{c2}_{ui})^2.
\end{equation}
Compared with concatenation and summation, our cross fusion module can more effectively capture the non-linearity between different types of features using a ``learn to fuse'' approach. Unlike the attention mechanism and meta paths methods, our fusion module are task-agnostic and generic, and can be applied to multiple tasks or scenarios. Moreover, our fusion module does not require additional learnable parameters and thus does not add complexity to the model training process. We further theoretically analyse the effectiveness of the proposed fusion method compared with the other fusion methods in Section~\ref{sec:analysis}.

\subsection{Model Training}\label{sec:training}
The model training can be divided into two stages: the first stage is to train the auxiliary feature extraction module and obtain the auxiliary feature, and the second stage is to train the graph feature extraction module and achieve the fusion of graph feature and auxiliary feature.

In the first stage, we use the following loss function to train the auxiliary feature extraction module:
\begin{equation}
    \mathcal{L}_{a}=\sum_{(u,i, r)\in\mathcal{O}} (r^a_{ui}-r)^2,\label{eq:loss_aux}
\end{equation}
where $\mathcal{O}$ is the set of all training triplets, $r$ is the true interaction score between user $u$ and item $i$. After the first training stage, we can obtain the user and item auxiliary features, then we incorporate the obtained auxiliary feature with user interactions to train the whole framework with the following loss in the second stage:
\begin{align}
    \mathcal{L}= \mathcal{L}_g+\lambda_1\mathcal{L}_{c1}+\lambda_2\mathcal{L}_{c2},\label{eq:loss}
\end{align}
where $\lambda_1$ and $\lambda_2$ are used to balance the importance of different terms in the loss. 

Algorithm~\ref{alg:algo} shows the training process of the framework in two stages. We first randomly initialize the model parameters (Line 1), then we train the auxiliary feature extraction module using user interactions and auxiliary information of users and items in the first stage (Line 2--Line 5). After training, we can obtain the trained user and item auxiliary features (Line 6), then we train the graph feature extraction module and achieve the fusion of graph feature and auxiliary feature simultaneously using the obtained user and item auxiliary features with user-item interactions in the second stage (Line 7--Line 9).

\begin{algorithm}[t!]
\caption{The training process of \ours}
\label{alg:algo}
\begin{flushleft}
\textbf{Input}: User-item interaction matrix $\mathbf{R}$, primary user auxiliary feature $\mathbf{X}$, primary item auxiliary feature $\mathbf{Y}$, graph feature extraction module parameters $\mathbf{\Theta}_g$, auxiliary feature extraction module parameters $\mathbf{\Theta}_a$, learning rates $\eta_1$ and $\eta_2$, number of training epoch $E$\\
\textbf{Output}: The trained model $\mathbf{\Theta}_g$ and $\mathbf{\Theta}_a$.\\
\end{flushleft}
\begin{algorithmic}[1]
\STATE Randomly initialize model parameters $\mathbf{\Theta}_g$, $\mathbf{\Theta}_a$.
\FOR{$e=1, 2, \cdots, E$}
\STATE Train auxiliary feature extraction module  using $\mathbf{R}$,  $\mathbf{X}$ and $\mathbf{Y}$ and obtain loss $\mathcal{L}_{a}$ through Equation (\ref{eq:loss_aux}).
\STATE Update auxiliary feature extraction module parameters by: $\Theta_a\leftarrow\Theta_a-\eta_1\cdot\frac{\partial\mathcal{L}_a}{\partial \Theta_a}$.
\ENDFOR
\STATE Obtain user and item auxiliary feature $\mathbf{A}_U$ and $\mathbf{A}_V$ from auxiliary feature extraction module.
\FOR{$e=1, 2, \cdots, E$}
\STATE Train graph feature extraction module using $\mathbf{R}$ , $\mathbf{A}_U$ and $\mathbf{A}_V$ and obtain loss $\mathcal{L}$ through Equation (\ref{eq:loss}). 
\STATE Update graph feature extraction module parameters by: $\Theta_g\leftarrow\Theta_g-\eta_2\cdot\frac{\partial\mathcal{L}}{\partial \Theta_g}$.
\ENDFOR
\end{algorithmic}
\end{algorithm}

\subsection{Inference}
In the inference stage, \ours takes as input graph feature to generate user embeddings and item embeddings to make recommendations, since auxiliary features of users and items have been encoded in the parameters learned by the graph feature extraction module through the back-propagation in the training. In the top-K recommendation task, for a given user $u$, we can infer which item $i$ that she/he will interact with by choosing the item $i'\in\mathcal{I}$ that has the maximum interaction score:
\begin{equation}
    i=\underset{i'\in\mathcal{I}}{\arg\max}~{\mathbf{g}_u^T\mathbf{g}}_{i'}.
\end{equation}

\section{Discussion}
In this section, we analyze the update direction of user and item embeddings for understanding the effectiveness of \ours from the perspective of gradient.  
Then, we discuss how to extend \ours in other feature fusion scenarios, e.g., the fusion of features in different time periods of
sequential recommendation models.

\subsection{Gradient Analysis}
\label{sec:analysis}
Firstly, we analyse how cross fusion module achieves the fusion of different types of features from the perspective of gradients. Equation~(\ref{eq:loss_aux}) shows the loss of auxiliary feature fusion module, in which the updating directions of user embedding $\mathbf{a}_u$ and item embedding $\mathbf{a}_i$ are as follows:
\begin{equation}
    \frac{\partial \mathcal{L}_a}{\partial \mathbf{a}_u}=\sum_{j\in\mathcal{N}(u)} w^{a1}_{u,j}{\mathbf{a}_j},\quad
    \frac{\partial \mathcal{L}_a}{\partial \mathbf{a}_i}=\sum_{v\in\mathcal{N}(i)}w^{a2}_{v,i}{\mathbf{a}_v},
\end{equation}
where $w^{a1}_{u,j}=2(\mathbf{a}_u^T\mathbf{a}_j-r)$ and $w^{a2}_{v,i}=2(\mathbf{a}_v^T\mathbf{a}_i-r)$. For the loss of graph feature extraction module, $\mathcal{L}_g$, we take mean square error (MSE) loss as an example in order to keep pace with $\mathcal{L}_a$,  while other losses, such as BPR loss~\cite{bpr}, can follow the same analysis and draw similar conclusions. $\mathcal{L}_g$ is defined as:
\begin{equation}
\mathcal{L}_g=\sum_{(u,i,r)\in\mathcal{O}}(\mathbf{g}_u^T\mathbf{g}_i-r)^2.
\end{equation}
Thus, the updating directions of user embedding $\mathbf{g}_u$ and item embedding $\mathbf{g}_i$ are as follows:
\begin{equation}
    \frac{\partial \mathcal{L}_g}{\partial \mathbf{g}_u}=\sum_{j\in\mathcal{N}(u)}w^{g1}_{u,j}\mathbf{g}_j,\quad
    \frac{\partial \mathcal{L}_g}{\partial \mathbf{g}_i}=\sum_{v\in\mathcal{N}(i)} w^{g2}_{v,i}\mathbf{g}_v,\label{eq:para1}
\end{equation}
where $w^{g1}_{u,j}=2(\mathbf{g}_u^T\mathbf{g}_j-r)$ and $w^{g2}_{v,i}=2(\mathbf{g}_v^T\mathbf{g}_i-r)$. Given the loss function of \ours defined in Equation~(\ref{eq:loss}), the updating direction of embeddings are:
\begin{equation}
    \frac{\partial \mathcal{L}}{\partial \mathbf{g}_u} = \sum_{j\in\mathcal{N}(u)}\left(w^{g1}_{u,j}\mathbf{g}_j+w^{c1}_{u,j}\mathbf{a}_j\right),\quad
    \frac{\partial \mathcal{L}}{\partial \mathbf{g}_i} = \sum_{v\in\mathcal{N}(i)}\left(w^{g2}_{v,i}\mathbf{g}_v+w^{c2}_{v,i}\mathbf{a}_v\right),\label{eq:eqn22}
\end{equation}
where $w^{c1}_{u,j}=2\lambda_1(\mathbf{g}_u^T\mathbf{a}_j-\mathbf{a}_u^T\mathbf{a}_j)$ and $w^{c2}_{v,i}=2\lambda_2(\mathbf{a}_v^T\mathbf{g}_i-\mathbf{a}_v^T\mathbf{a}_i)$. 

Comparing the updating directions of user embedding $\mathbf{g}_u$ in Equation~(\ref{eq:para1}) and Equation~(\ref{eq:eqn22}), we find that the updating direction of $\mathbf{g}_u$ in Equation~(\ref{eq:eqn22}) is not only based on the item embedding $\mathbf{g}_j$, but also based on the item embedding $\mathbf{a}_j$. $\mathbf{a}_j$ is trained in the first stage as mentioned in Section~\ref{sec:training} which has sufficiently encoded the auxiliary feature after training, and it is fixed during training $\mathbf{g}_u$ in Equation~(\ref{eq:eqn22}). Thus, \ours comprehensively considers two types of features, i.e., graph feature (${\mathbf{g}_j}$) and auxiliary feature (${\mathbf{a}_j}$) to update the embedding, therefore it can achieve the fusion of different features. Similar findings can be obtained by comparing the updating directions of item embedding $\mathbf{g}_i$ in Equation~(\ref{eq:para1}) and Equation~(\ref{eq:eqn22}).

We can also conclude that \ours is superior to other fusion methods, e.g., concatenation and (weighted) summation. The loss of the concatenation based method can be represented as:
\begin{equation}
\mathcal{L}=\sum_{(u,i,r)\in\mathcal{O}}(\mathbf{a}_u^T\mathbf{a}_i+\mathbf{g}_u^T\mathbf{g}_i-r)^2,
\end{equation}
and the updating processes of embeddings are: 
\begin{equation}
        \frac{\partial \mathcal{L}}{\partial \mathbf{g}_u}=\sum_{j\in\mathcal{N}(u)} 2(\mathbf{a}_u^T\mathbf{a}_j+\mathbf{g}_u^T\mathbf{g}_j-r)\mathbf{g}_j,\quad
        \frac{\partial \mathcal{L}}{\partial \mathbf{g}_i}=\sum_{v\in\mathcal{N}(i)} 2(\mathbf{a}_v^T\mathbf{a}_i+\mathbf{g}_v^T\mathbf{g}_i-r)\mathbf{g}_v.\label{eq:concat}
\end{equation}
Comparing the update directions of $\mathbf{g}_u$ in Equation (\ref{eq:eqn22}) and Equation (\ref{eq:concat}), the former considers both item graph feature $\mathbf{g}_j$ and item auxiliary feature $\mathbf{a}_j$, while the latter only considers the item graph feature $\mathbf{g}_j$. This shows the superiority of our fusion module that can achieve more effective fusion than concatenation methods.

The loss of the (weighted) summation based fusion methods can be described as:
\begin{equation}
\mathcal{L}=\sum_{(u,i,r)\in\mathcal{O}}\left((\mathbf{W}_1\mathbf{a}_u+\mathbf{W}_2\mathbf{g}_u)^T(\mathbf{W}_3\mathbf{a}_i+\mathbf{W}_4\mathbf{g}_i)-r\right)^2,
\end{equation}
where $\mathbf{W}_1$, $\mathbf{W}_2$, $\mathbf{W}_3$, and $\mathbf{W}_4$ are weights, and the updating processes of embeddings are:
\begin{equation}
    \begin{aligned}
        \frac{\partial \mathcal{L}}{\partial \mathbf{g}_u}&=\sum_{j\in\mathcal{N}(u)} 2\left((\mathbf{W}_1\mathbf{a}_u+\mathbf{W}_2\mathbf{g}_u)^T(\mathbf{W}_3\mathbf{a}_j+\mathbf{W}_4\mathbf{g}_j)-r\right){\mathbf{W}_2}^T(\mathbf{W}_3\mathbf{a}_j+\mathbf{W}_4\mathbf{g}_j),\\
        \frac{\partial \mathcal{L}}{\partial \mathbf{g}_i}&=\sum_{v\in\mathcal{N}(i)} 2\left((\mathbf{W}_1\mathbf{a}_v+\mathbf{W}_2\mathbf{g}_v)^T(\mathbf{W}_3\mathbf{a}_i+\mathbf{W}_4\mathbf{g}_i)-r\right){\mathbf{W}_4}^T(\mathbf{W}_1\mathbf{a}_v+\mathbf{W}_2\mathbf{g}_v).\label{eq:summation}
    \end{aligned}
\end{equation}
Though both the updating directions of $\mathbf{g}_u$ in Equation (\ref{eq:eqn22}) and Equation (\ref{eq:summation}) consider the item embeddings $\mathbf{a}_j$ and $\mathbf{g}_j$, the updating direction of the former is more flexible since the ratio between $\mathbf{a}_j$ and $\mathbf{g}_j$ varies for different user-item pairs while that in the updating direction of the latter is fixed, thus the former has a more flexible updating direction than the latter. Therefore, $\mathbf{g}_u$ in Equation (\ref{eq:eqn22}) can achieve  personalized updating while $\mathbf{g}_u$ in Equation (\ref{eq:summation}) cannot.

The over-smoothing problem in GNN-based collaborative filtering methods illustrates that as the number of layers in GNN increases, the user and item features become increasingly similar, which could result in similar update directions of user and item, i.e., $\sum_{j\in\mathcal{N}(u)}w^{g1}_{u,j}{\mathbf{g}_j}$ and $\sum_{v\in\mathcal{N}(i)} w^{g2}_{v,i}{\mathbf{g}_v}$ and thus similar node representations of user $u$ and item $i$. However, by incorporating  auxiliary features, the updating directions are no longer similar to those before, since the newly added direction $\sum_{j\in\mathcal{N}(u)}w^{c1}_{u,j}{\mathbf{a}_j}$ and $\sum_{v\in\mathcal{N}(i)}w^{c2}_{v,i}{\mathbf{a}_v}$ are generated from different feature spaces that are not linear to each other. Therefore, the over-smoothing issue can be alleviated by introducing auxiliary features. It is worth noting that the effectiveness of this alleviation depends on the informativeness of the auxiliary features, as more informative ones can provide a superior solution.

\subsection{Extension of \ours}\label{sec:extension}

\ours has a high level of universality, as it is capable of fusing a wide range of features, including the fusion of graph features from user interactions and text features from user reviews, the fusion of features from different time periods in sequential recommendation models, and the fusion of features extracted by various recommendation models. As an illustration, we utilize the fusion of features from different time periods as an example to demonstrate the applicability of \ours in other feature fusion tasks.

In sequential recommendation, a standard approach is to divide user interactions into several time periods in chronological order and then apply sequential models to extract the temporal features of the users and items changing over time. Suppose that the temporal features of user $u$ and item $i$ in time period $t$ are $\mathbf{h}_u^{t}$ and $\mathbf{h}_i^{t}$ respectively, then the temporal features of user $u$ and item $i$ in time periods before $t$ (denoted as $t_u^-$ and $t_i^-$) can be used as the auxiliary feature to help the feature learning in time period $t$. More specifically, the learning process can be described as minimizing the following loss function:
\begin{equation}
    \mathcal{L}=
    \mathcal{L}_t+\sum_{(u,i, t)\in\mathcal{O}}\lambda_1({\mathbf{h}_u^{t_u^-}}^T\mathbf{h}_i^{t_u^-}- {\mathbf{h}_u^{t}}^T\mathbf{h}_i^{t_u^-})^2+\lambda_2({\mathbf{h}_u^{t_i^-}}^T\mathbf{h}_i^{t_i^-}- {\mathbf{h}_u^{t_i^-}}^T\mathbf{h}_i^{t})^2,
\end{equation}
where $\mathcal{L}_t$ is the loss function of existing sequential recommendation models, such as RRN~\cite{rrn} and JODIE~\cite{jodie}, and $t_u^-$ and $t_i^-$ are the time periods before $t$ during which user $u$ and item $i$ have interactions, respectively. $\lambda_1$ and $\lambda_2$ are the coefficients to balance the losses. The two newly added terms act as a cross fusion mechanism, smoothing the changes in representations over time. Since the updating direction of $\mathbf{h}_u^{t}$ (or $\mathbf{h}_i^{t}$) takes $\mathbf{h}_j^{t_u^-} (j\in\mathcal{N}(u))$ (or $\mathbf{h}_v^{t_i^-} (v\in\mathcal{N}(i))$) into consideration, the training of the new model can benefit from the previous models, 
thus improving the performance of recommendation algorithms.

\section{Experiments}
\subsection{Experimental Setup}
\begin{table*}[t!]
\renewcommand{\arraystretch}{1.3}
\caption{The statistics of the six datasets used in our experiment to evaluate the effectiveness of \ours.}
\label{tab:dataset}
\centering
\resizebox{1.0\linewidth}{!}{
\begin{tabular}{l|cccccc}
\toprule
& \bfseries \# users &\bfseries \# items &\bfseries \# records & \bfseries Density & \bfseries User side information& \bfseries Item side information \\
\midrule
Post&500&6,000&71,800&0.0239&gender, city, academics&category\\
ML1M &6,040&3,706&1,000,208&0.0447&age, gender, occupation&category\\
Music &4,999&15,000&2,934,797&0.0391&city, bd, gender, registered\_via&genre\_ids, language\\
\midrule
& \bfseries \# users &\bfseries \# items &\bfseries \# records & \bfseries Density & -- & -- \\
\midrule
Book&6,936&4,926&207,433&0.0061&--&--\\
Yelp&4,989&4,987&286,746&0.0115&--&--\\
\midrule
& \bfseries \# users &\bfseries \# items &\bfseries \# records & \bfseries Density & \bfseries Training time span&\bfseries Test time span \\
\midrule
Video&4,922&9,816&115,966&0.0024&2010.01-2016.12&2017.01-2018.10\\
ML1M&5,620&3,673&827,373&0.0401&2000.06-2000.11&2000.12\\
\bottomrule
\end{tabular}
}
\end{table*}
\subsubsection{Datasets}\label{chp:base}

We adopt six public datasets to evaluate the performance of the proposed \ours framework: (1) {\bf Post} is a user's post view dataset, collected by Kaggle user, that contains both user interactions and the auxiliary features of users and items; (2) {\bf ML1M (MovieLens 1M)} is a movie recommendation dataset that is collected by GroupLens Research from the MovieLens web site; (3) {\bf Music (KKBox Music)} is a music dataset that is released by KKBOX in the WSDM 2018 Challenge; (4) {\bf Book (Amazon Book)} and {\bf Video (Amazon Video Games)} are two product recommendation datasets released by Amazon.com; (5) {\bf Yelp} is a review dataset about the restaurant and food released by Yelp. 

The six datasets are used to verify the fusion capability of \ours in four different scenarios, including the fusion of auxiliary feature and graph feature (Post, ML1M and Music), the fusion of graph feature and text feature (Book and Yelp), the fusion of features in different time periods of sequential recommendation models (Video and ML1M) and the fusion of features extracted by different models (ML1M and Music). 
All datasets are divided into training and test sets by the ratio of 80\%:20\%, and we take 10\% of the training set as the validation set to tune the hyper-parameters. Table~\ref{tab:dataset} describes the detailed statistics of the six datasets.

\subsubsection{Compared Methods}
We evaluate the effectiveness of \ours by comparing with the following eleven collaborative filtering methods in the four different scenarios mentioned above. Table~\ref{tab:baselines} summarizes the application of the eleven compared methods in four different scenarios.
\begin{itemize}
    \item GIN~\cite{gin} is a powerful structural feature extraction algorithm based on GNN whose expressive ability is proved to be as powerful as Weisfeiler-Lehman graph isomorphism test by adopting sum aggregation and multi-layer perceptrons. 
    \item LightGCN~\cite{lightgcn} is a GCN-based CF algorithm which simplifies the structure of GCN by removing feature transformation and nonlinear activation to improve the performance.
    \item LR-GCCF~\cite{lrgccf} is a linear method, which removes non-linearities and adds a residual network structure, that can be scaled to large datasets and achieve better efficiency and effectiveness.
    \item IMP-GCN~\cite{impgcn} is a GCN-based method that takes high-order neighboring users' interests into account to avoid propagating negative information from high-order neighbors into user embedding learning and alleviate over-smoothing issue.
    \item HAN~\cite{han} is a GNN-based method that learns node embedding through meta paths and two attention mechanisms (node-level attention and semantic-level attention) hierarchically. 
    \item AM-GCN~\cite{amgcn} is a GNN-based method that learns node representations by using the attention mechanism to fuse node feature, topological structures and their combinations.
    \item GraphRec~\cite{graphrec} is a GNN-based method that takes both the users’ social relationship and the user’s interactions into account to learn user and item features.
    \item RRN~\cite{rrn} is a sequential recommendation method that uses two RNNs to model users’ and items’ temporal features respectively. 
    \item GCMC~\cite{gcmc} is a graph based matrix completion method that uses a graph encoder to learn user and item representations and a bi-linear decoder to predict the interactions.
    \item GCMCRNN~\cite{gcmcrnn} is a sequential CF method that uses GCMC to learn spatial features from user-item interaction graphs in all time periods and then uses RNN to extract temporal features.
    \item JODIE~\cite{jodie} is a graph based sequential recommendation method that employs two RNNs to learn the embedding trajectories of users and items from user-item interaction graph.    
\end{itemize}

We also conduct an ablation study to verify the effectiveness of \ours in feature fusion by comparing with the following feature fusion methods:
\begin{itemize}
    \item Concatenation. In this setting, we design a variant that replaces our cross fusion module with a simple concatenation operation.
    \item Summation. In this setting, we design two variants that replace the cross fusion module with two summations, plain summation that directly adds different features and weighted summation that adds different features with learnable weights.
    \item Attention. In this setting, we design a variant that replaces the cross fusion module with the attention mechanism proposed in AM-GCN~\cite{amgcn}. 
    \item Meta-path. In this setting, we compare with HAN~\cite{han} to demonstrate the effectiveness of the proposed cross fusion module compared against manually designed meta paths.
\end{itemize}

\begin{table*}[t!]
\caption{The application of the eleven collaborative filtering methods in four different scenarios: (a) the fusion of auxiliary feature and graph feature, (b) the fusion of graph feature and text feature, (c) the fusion of features in different time periods of sequential recommendation models, and (d) the fusion of features extracted by different graph neural network models.}
\label{tab:baselines}
\centering
\resizebox{1.0\linewidth}{!}
{
\begin{tabular}{l|ccccccccccc}
\toprule
& \bfseries GIN &\bfseries LightGCN &\bfseries LR-GCCF & \bfseries IMP-GCN & \bfseries HAN & \bfseries AM-GCN & \bfseries GraphRec& \bfseries RRN  & \bfseries GCMCRNN & \bfseries JODIE & \bfseries GCMC\\
\midrule
(a)&\checkmark&\checkmark&\checkmark&\checkmark&\checkmark&\checkmark&\checkmark&&&&\\
(b)&\checkmark&\checkmark&&\checkmark&&&&&&&\\
(c)&&&&&&&&\checkmark&\checkmark&\checkmark&\\
(d)&\checkmark&\checkmark&&\checkmark&&&&&&&\checkmark\\
\bottomrule
\end{tabular}
}
\end{table*}

\subsubsection{Metrics}
We evaluate \ours and the compared recommendation algorithms in the Top-N recommendation task with three ranking metrics: 1) F1, which is the harmonic mean of precision and recall; 2) Mean Reciprocal Rank (MRR), which evaluates the performance by the ranking of the first right recommendation in the list; and 3) Normalized Discounted Cumulative Gain (NDCG), which evaluates the gap between recommended item lists and ground-truth item lists. For all metrics, we report their results when N=5 and N=10 in the experiments.

\subsubsection{Implementation Details}
We implement \ours  with Pytorch\footnote{https://pytorch.org} and Tensorflow\footnote{https://www.tensorflow.org/}. We also use DGL\footnote{https://docs.dgl.ai}, a Python library for easy implementation of graph neural networks, to implement GCNs in the auxiliary feature extraction module. We train the model 200 epochs. In the graph feature extraction module, we tune the hyper-parameters of graph embedding methods according to corresponding papers. In the auxiliary feature extraction module, we tune the user/item threshold $\epsilon$ in [0.1, 0.3, 0.5, 0.7, 0.9]. In the cross fusion module, we tune the weights $\lambda_1$ and $\lambda_2$ in [0.001, 0.01, 0.05, 0.1, 1], and search the number of layers in GCN in [2, 3, 4, 5]. For two learning rates $\eta_1$ and $\eta_2$ in Algorithm~\ref{alg:algo}, we tune them in [0.0003, 0.001, 0.003, 0.01, 0.03]. For all baselines, we directly use their released code and tune the hyper-parameters according to their papers.

\begin{table*}[!t]
\renewcommand{\arraystretch}{1.3}
\caption{The results of accuracy comparison. Since HAN-2 occurred the Out-Of-Memory problem on Music dataset, so we do not report the results.}
\label{tab:acc_com}
\centering
\resizebox{1.0\linewidth}{!}{
\begin{tabular}{l|ccccccccc}
\toprule
 &\multicolumn{3}{c}{Post} & \multicolumn{3}{c}{ML1M}& \multicolumn{3}{c}{Music}\\
\cmidrule(r){2-4}\cmidrule(r){5-7}\cmidrule(r){8-10}
&F1@5&MRR@5&NDCG@5&F1@5&MRR@5&NDCG@5&F1@5&MRR@5&NDCG@5\\
\midrule
HAN-1&0.0017&0.0129&0.0168&0.0118&0.0778&0.0893&0.0076&0.1038&0.1299\\
HAN-2&0.0019&0.0129&0.0168&0.0125&0.0833&0.0950&--&--&--\\
AM-GCN&0.0025&0.0120&0.0168&0.0106&0.0763&0.0856&0.0061&0.1626&0.1675\\
GraphRec&0.0021&0.0144&0.0187&0.0059&0.0364&0.0461&0.0069&0.1562&0.1931\\
\midrule
GIN&0.0023&0.0134&0.0193&0.0720&0.2720&0.3153&0.0619&0.5202&0.5744\\
\ours-GIN&0.0029&0.0177&0.0225&0.0746&0.2880&0.3294&0.0635&0.5264&0.5827\\
Impr.(\%)&+26.09&+32.09&+16.58&+3.61&+5.88&+4.47&+2.58&+1.19&+1.44\\
\midrule
LR-GCCF&0.0016&0.0151&0.0180&0.0683&0.2791&0.3183&0.0634&0.5249&0.5808\\
\ours-LR-GCCF&0.0018&0.0152&0.0206&0.0751&0.2927&0.3337&0.0653&0.5419&0.5932\\
Impr.(\%)&+12.50&+0.66&+14.44&+9.96&+4.87&+4.84&+3.00&+3.24&+2.13\\
\midrule
LightGCN&0.0026&0.0123&0.0180&0.0626&0.2872&0.3260&0.0423&0.3782&0.4231\\
\ours-LightGCN&0.0027&0.0156&0.0214&0.0814&0.3072&0.3512&0.0632&0.5224&0.5781\\
Impr.(\%)&+3.85&+26.83&+18.89&+30.03&+6.96&+7.73&+49.41&+38.13&+36.63\\
\midrule
IMP-GCN&0.0026&0.0145&0.0176&0.0828&0.2886&0.3375&0.0636&0.5227&0.5789\\
\ours-IMP-GCN&0.0037&0.0157&0.0203&0.0819&0.3078&0.3519&0.0645&0.5293&0.5858\\
Impr.(\%)&+42.31&+8.28&+15.34&-1.09&+6.65&+4.27&+1.42&+1.26&+1.19\\
\midrule\midrule
 &\multicolumn{3}{c}{Post} & \multicolumn{3}{c}{ML1M}& \multicolumn{3}{c}{Music}\\
\cmidrule(r){2-4}\cmidrule(r){5-7}\cmidrule(r){8-10}
&F1@10&MRR@10&NDCG@10&F1@5&MRR@10&NDCG@10&F1@10&MRR@10&NDCG@10\\
\midrule
HAN-1&0.0021&0.0126&0.0193&0.0178&0.1018&0.1201&0.0115&0.1103&0.1572\\
HAN-2&0.0025&0.0160&0.0231&0.0188&0.1088&0.1272&--&--&--\\
AM-GCN&0.0034&0.0131&0.0238&0.0196&0.0812&0.1077&0.0075&0.1758&0.1888\\
GraphRec&0.0032&0.0180&0.0265&0.0131&0.552&0.0780&0.0125&0.1856&0.2465\\
\midrule
GIN&0.0033&0.0153&0.0266&0.1015&0.2627&0.3364&0.0956&0.4849&0.5594\\
\ours-GIN&0.0035&0.0192&0.0286&0.1062&0.2992&0.3639&0.0991&0.5023&0.5775\\
Impr.(\%)&+6.06&+25.49&+7.52&+4.63&+13.89&+8.17&+3.66&+3.59&+3.24\\
\midrule
LR-GCCF&0.0026&0.0158&0.0241&0.0968&0.2893&0.3496&0.0977&0.5027&0.5759\\
\ours-LR-GCCF&0.0027&0.0187&0.0283&0.1044&0.3037&0.3661&0.1000&0.5116&0.5842\\
Impr.(\%)&+3.85&+18.35&+17.43&+7.85&+4.98&+4.72&+2.35&+1.77&+1.44\\
\midrule
LightGCN&0.0038&0.0166&0.0278&0.0854&0.2912&0.3506&0.0638&0.3625&0.4302\\
\ours-LightGCN&0.0039&0.0188&0.0288&0.1144&0.3039&0.3746&0.0978&0.5015&0.5750\\
Impr.(\%)&+2.63&+13.25&+3.60&+33.96&+4.36&+6.85&+53.29&+38.34&+33.66\\
\midrule
IMP-GCN&0.0032&0.0142&0.0216&0.1081&0.2712&0.3483&0.0972&0.4888&0.5661\\
\ours-IMP-GCN&0.0036&0.0142&0.0220&0.1140&0.3121&0.3790&0.0998&0.5063&0.5811\\
Impr.(\%)&+12.50&+0.00&+1.85&+5.46&+15.08&+8.81&+2.67&+3.58&+2.65\\
\bottomrule
\end{tabular}
}
\end{table*}

\subsection{Accuracy Comparison}
In this experiment, we choose HAN, AM-GCN, GraphRec, GIN, LR-GCCF, LightGCN and IMP-GCN as baselines. HAN and AM-GCN can capture graph feature from user interactions and auxiliary feature from user and item auxiliary information simultaneously, and fuse them using meta path and attention mechanism respectively. For HAN, we use the two variants HAN-1 and HAN-2 in our experiments, the difference between them is the meta paths that they adopt. In HAN-1, we use all 1-st order meta paths (e.g., user-city-user for Post), and in HAN-2, besides 1-st order meta paths, we add partial 2-nd order meta paths (e.g., user-city-user-academics-user for Post) since the amount of 2-nd order meta paths is large. For GraphRec, it was originally proposed to fuse user interactions and user social relationships. However, user social relationships are missing in the datasets, thus we use user-user interaction graph $\mathbf{U}$ instead of user social relationship graph in the social aggregation of GraphRec. GIN, LR-GCCF, LightGCN and IMP-GCN can only extract structural features from user-item interactions. Table~\ref{tab:acc_com} shows the accuracy comparison results, from which we have the following observations.
\begin{itemize}
    \item {\bf Comparison to meta-path-based method.} HAN-1 and HAN-2 achieve relatively lower performance in most cases, which shows that meta paths may not the optimal method to fuse graph feature and auxiliary feature. This is because they rely on manually designed meta paths, which may lead to inefficiencies in the fusion process due to limited human knowledge on the fusion task. 
    \item {\bf Comparison to attention-based method.} Though AM-GCN has better performance than HAN in most cases, it underperforms \ours. AM-GCN  proposed to fuse topological structure and node feature using attention mechanism for node classification in traditional graph learning task, which may not be unsuitable for recommendation tasks where data is usually very sparse. The introduction of additional parameters also increases the difficulty of model training, thus resulting in inferior performance.
    \item {\bf Comparison to concatenation-based method.} GraphRec achieves better performance than AM-GCN in most cases, this is because it takes user social relationships into consideration. However, it performs worse than \ours, which we attribute to a lack of feature relation modeling and an incomplete representation of user social relationships, as user social relationships and user-user interaction graphs may significantly differ.
    \item {\bf Fusion performance on existing GNN-based CF methods.} The performance of GIN, LR-GCCF, LightGCN and IMP-GCN has consistently improved when equipped with \ours. The reason is that auxiliary feature, as a supplement to user interactions, can significantly improve the prediction accuracy of the CF models. With the help of the proposed cross fusion module, \ours can transfer knowledge from auxiliary feature extraction module to GIN, LR-GCCF, LightGCN, and IMP-GCN, thus effectively fusing these two types of features and improving the performance. 
\end{itemize}

\begin{figure*}[!h]
\centering
\includegraphics[width=1.0\textwidth]{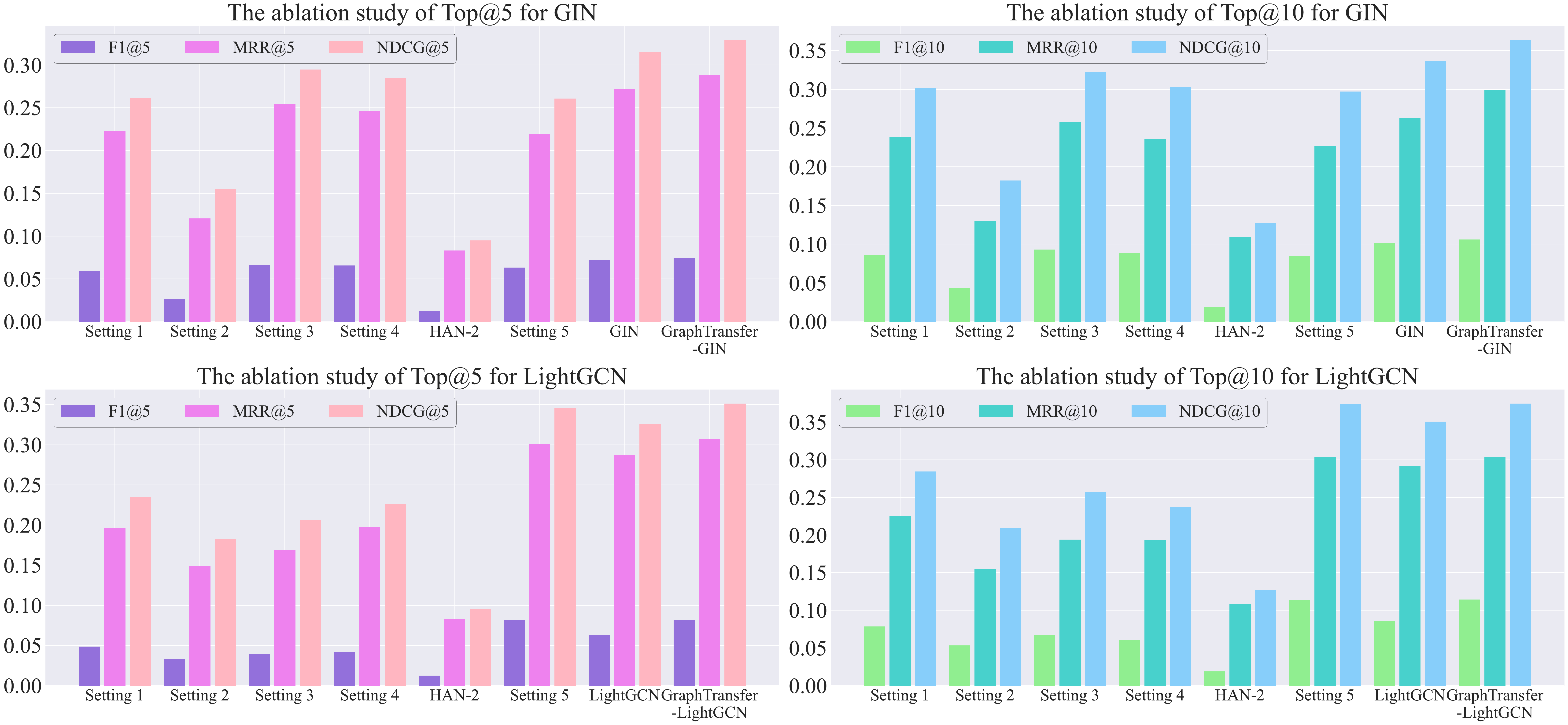}
\caption{The ablation study results for \ours-GIN (the upper two figures) and \ours-LightGCN (the lower two figures) under eight different settings on ML1M dataset. Setting 1--Setting 4 represent replacing cross fusion module in \ours with concatenation, plain summation, weighted summation and attention mechanism respectively. Setting 5 represents removing user and item GCNs in \ours.}
\label{fig:abl_stu}
\end{figure*}

\subsection{Ablation Study}
We analyze the impacts of the auxiliary feature extraction module and cross fusion module on the performance of \ours through the ablation study. Figure~\ref{fig:abl_stu} shows the results of the ablation study with respect to GIN (the upper two figures) and LightGCN (the lower two figures) under eight settings. In Setting 1--4, we replace cross fusion module in \ours with concatenation, plain summation, weighted summation and attention respectively, and in Setting 5 we remove user and item GCNs in auxiliary feature extraction module. All experiments are conducted on the ML1M dataset.

From the results presented in Figure~\ref{fig:abl_stu}, we can see that the trends observed in both GIN and LightGCN are similar. In order to simplify the discussion, we will focus on the results of GIN as a representative example. Specifically, we have the following observations from the results.
\begin{itemize}
    \item Comparing between Setting 1--4 and \ours-GIN, we find that \ours-GIN achieves the best results in all cases. This is because concatenation based method (Setting 1) and plain summation based methods (Setting 2), which directly concatenate and add different features in the fusion process, are too simple to effectively capture the non-linearity among different types of features. Though learnable summation (Setting 3) and attention (Setting 4) outperform concatenation and plain summation in feature fusion, they still underperform \ours-GIN. This is mainly due to the introduction of additional learnable parameters, which increases not only the expressivity of model but also complexity and difficulty of training. The task-specific attention mechanism, which is borrowed from AM-GCN, also hinders the performance of the model. The meta-paths-based method (HAN-2) relies on specialized domain knowledge, specifically, manually designed meta paths, for feature fusion. As a result, its performance is highly dependent on the quantity and quality of these meta paths used. However, our proposed cross fusion module (\ours) utilizes a unique cross-dot-product approach for fusing different types of features. Our method is able to capture the non-linearity between different features, while also simplifying the model training process by not introducing additional parameters. These characteristics lead to superior results as shown in the experiment.
    
    \item Comparing Setting 5 with \ours-GIN, the latter still achieves superior results, which proves the constructed user-user interaction graph and item-item interaction graph are beneficial to extract auxiliary feature. User-user interaction graph (or item-item interaction graph) constructs the relationship between users (or items) that have similar characteristics. Through graph convolutions on these two graphs, the representation learning of users and items can be enhanced and the user and item features can become more distinctive, thus obtaining more promising results.

    \item \ours-GIN achieves better accuracy than standard GIN, which demonstrates the effectiveness of fusing graph feature and auxiliary feature in improving the performance of recommendation algorithms. Compared with graph features that describe users and items from their preference, auxiliary features depict users and items from a different perspective, i.e., the essence of them, which are orthogonal and complementary, thus integrating them can make the modeling of user and item more accurate. Through our proposed cross fusion module, \ours-GIN fuses these two types of feature effectively, thus achieving superior performance.
\end{itemize}

\begin{figure*}[!h]
\centering
\includegraphics[width=1.0\textwidth]{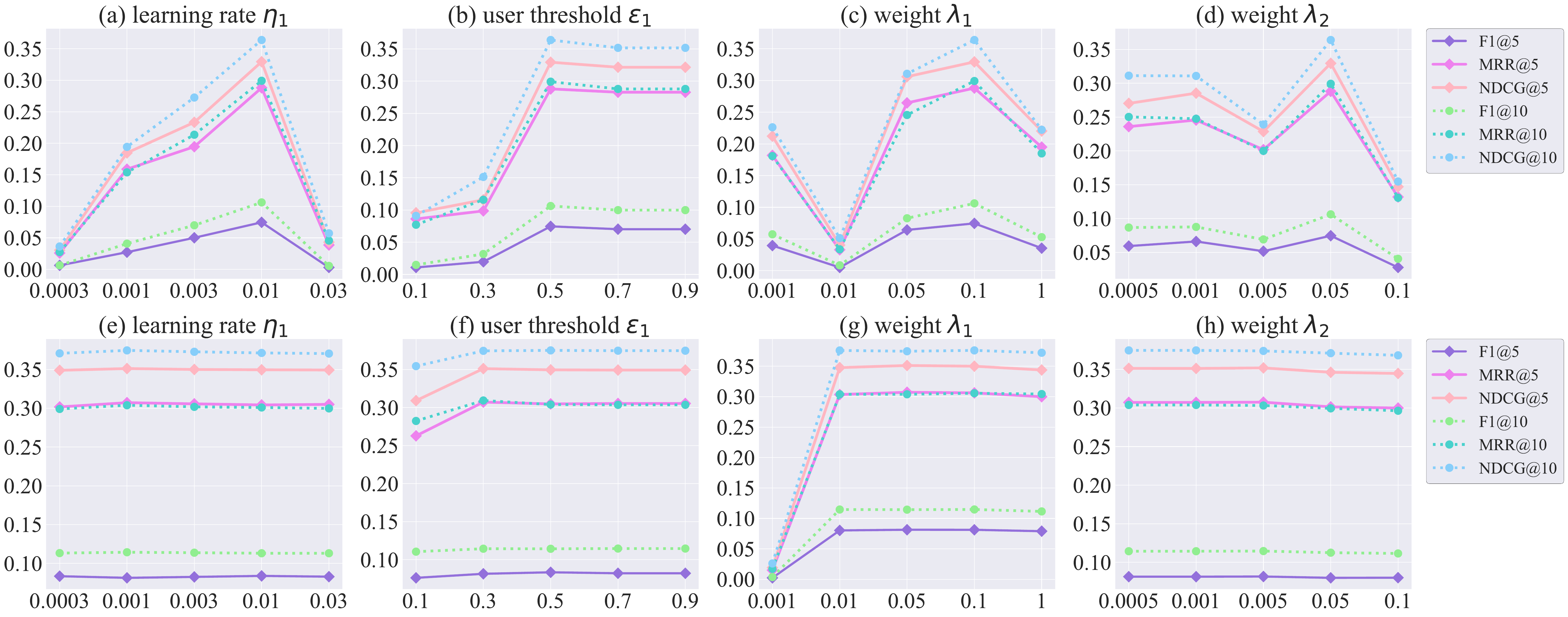}
\caption{The sensitivity analysis results with four hyper-parameters for \ours-GIN ((a)--(d)) and \ours-LightGCN ((e)--(h)) on ML1M: learning rate of auxiliary feature extraction module $\eta_1$, user threshold $\epsilon_1$, weight $\lambda_1$ and $\lambda_2$ in Equation~(\ref{eq:loss}), respectively.}
\label{fig:hps_ana}
\end{figure*}

\subsection{Sensitivity Analysis}
In this experiment, we analyse how \ours performs with four important hyper-parameters:  learning rate $\eta_1$ of auxiliary feature extraction module (in Algorithm~\ref{alg:algo}), user threshold $\epsilon_1$ and the weights $\lambda_1$ and $\lambda_2$ to balance different components of cross fusion module in Equation~(\ref{eq:loss}). The experiments are conducted on ML1M dataset. Figure~\ref{fig:hps_ana} shows the performance of \ours-GIN (the upper four figures (a)--(d)) and \ours-LightGCN (the lower four figures (e)--(h)) with respect to four different hyper-parameters.

\subsubsection{Impacts of learning rate $\eta_1$ of auxiliary feature extraction module}
Figure~\ref{fig:hps_ana} (a) shows the performance of \ours-GIN with learning rate $\eta_1$, varying in [0.0003, 0.001, 0.003, 0.01, 0.03]. We can observe that \ours-GIN achieves the best performance when $\eta_1=0.01$. It makes sense that with the increase of $\eta_1$, the model can converge to the local optimum faster and achieve the (locally) best performance. However, if $\eta_1$ is too large, the model will oscillate back and forth around the local optimum, or even deviate from the local optimum.

\subsubsection{Impacts of user threshold $\epsilon_1$}
Figure~\ref{fig:hps_ana} (b) shows the performance of \ours-GIN when user threshold $\epsilon_1$ varies in [0.1, 0.3, 0.5, 0.7, 0.9]. We can conclude from the results that \ours-GIN achieves best performance at $\epsilon_1=0.5$. As the value of $\epsilon_1$ becomes too small, the majority of users tend to have many connections or become closely connected in the user-user interaction graph, creating an excessive amount of noise and making it difficult for graph neural networks to extract useful features for prediction. Conversely, when $\epsilon_1$ is too large, there are few connections among users, leading to a sparse graph that limits the graph neural network's ability to collect and use neighboring information, resulting in poor performance. Therefore, a suitable value of $\epsilon_1$ is crucial in constructing an effective user-user interaction graph, allowing graph neural networks to extract relevant and informative features that accurately capture the characteristics of users and items.

\subsubsection{Impacts of weights $\lambda_1$ and $\lambda_2$}
Figure~\ref{fig:hps_ana} (c) shows the performance of \ours-GIN with respect to $\lambda_1$ varying in [0.001, 0.01, 0.05, 0.1, 1]. As the value of $\lambda_1$ increases, the performance of the model improves. This is due to the incorporation of additional knowledge from the auxiliary feature extraction module into \ours-GIN, leading to increased accuracy. The optimal performance of \ours-GIN is achieved when  $\lambda_1 = 0.1$. However, when $\lambda_1 > 0.1$, there is a decrease in accuracy as the model becomes too heavily reliant on the auxiliary feature, neglecting the most crucial graph features. Similar trends can be observed in Figure~\ref{fig:hps_ana} (d) where the performance peaks at $\lambda_2=0.05$ due to the same reason. 

For \ours-LightGCN, similar findings can be found in the lower four figures of Figure~\ref{fig:hps_ana}, where the best hyperparameters are: learning rate $\eta_1=0.001$, user threshold $\epsilon_1=0.3$, weight $\lambda_1=0.05$ and weight $\lambda_2=0.001$.

\subsection{Case Study}
We take LightGCN as an example to illustrate that, after the introduction of auxiliary feature through \ours, the category distribution of items recommended to each user has a higher consistency with that of user's historical preference (according to the user's historical interactions) than that without auxiliary feature. The experiment was conducted on ML1M. The category distribution of user $u$'s historical preference $p_u$ is defined as:
\begin{equation}
    p_{u}(\text{category}=l)=\frac{C_{ul}}{\sum_{k=1}^LC_{uk}},
\end{equation}
where $C_{ul}~(u=1,2,\cdots,N, l=1,2,\cdots, L) $ represents the number of appearances of the $L$-th item category in the user $u$'s historical items, $N$ represents the number of users, and $L$ represents the number of categories. Similarly, we can calculate the preference distributions of the predicted items for user $u$ from LightGCN and GraphTranfer-LightGCN as $q^{(1)}_u$ and $q^{(2)}_u$, respectively. Then the Kullback–Leibler (KL) divergence can be used to evaluate the quality of the predicted preference distribution, where larger KL divergence indicates smaller similarity. For instance, if $\mathbf{KL}(p, q^{(1)}) > \mathbf{KL}(p, q^{(2)})$, it indicates that it is useful to introduce auxiliary feature that makes the predicted preference distribution with a higher consistency with that of users' historical preference. The KL divergence between $p$ and $q$ is:
\begin{equation}
    \mathbf{KL}(p,q) = \frac{1}{N}\sum_{u=1}^N\sum_{l=1}^Lp_u(l)\ln\frac{p_u(l)}{q_u(l)}.
\end{equation}
We tend to focus on the most common $K$ categories of items in our experiment, while less common categories may be considered noisy, as a movie may be associated with multiple categories in the MovieLens dataset and a user's viewing behavior may not necessarily reflect an interest in all of those categories.

\begin{table}[t!]
\renewcommand{\arraystretch}{1.3}
\caption{The KL divergence between the item category distributions in the recommended item lists and the actual user histories in  LightGCN and \ours-LightGCN. Lower values indicate better match with groundtruth.}
\label{tab:klvalue}
\centering
\small
\begin{tabular}{l|ccccc}
\toprule
\bfseries Model& \bfseries $K = 6$ & \bfseries $K = 8$ & \bfseries $K = 10$ &\bfseries $K = 12$ & \bfseries $K = 14$\\
\midrule
LightGCN&0.0163&0.0104&0.0075&0.0058&0.0048\\
\makecell[c]{\ours-LightGCN}&0.0139&0.0093&0.0070&0.0056&0.0047\\
\bottomrule
\end{tabular}
\end{table}

\begin{figure}[!h]
\centering
\includegraphics[width=0.8\linewidth]{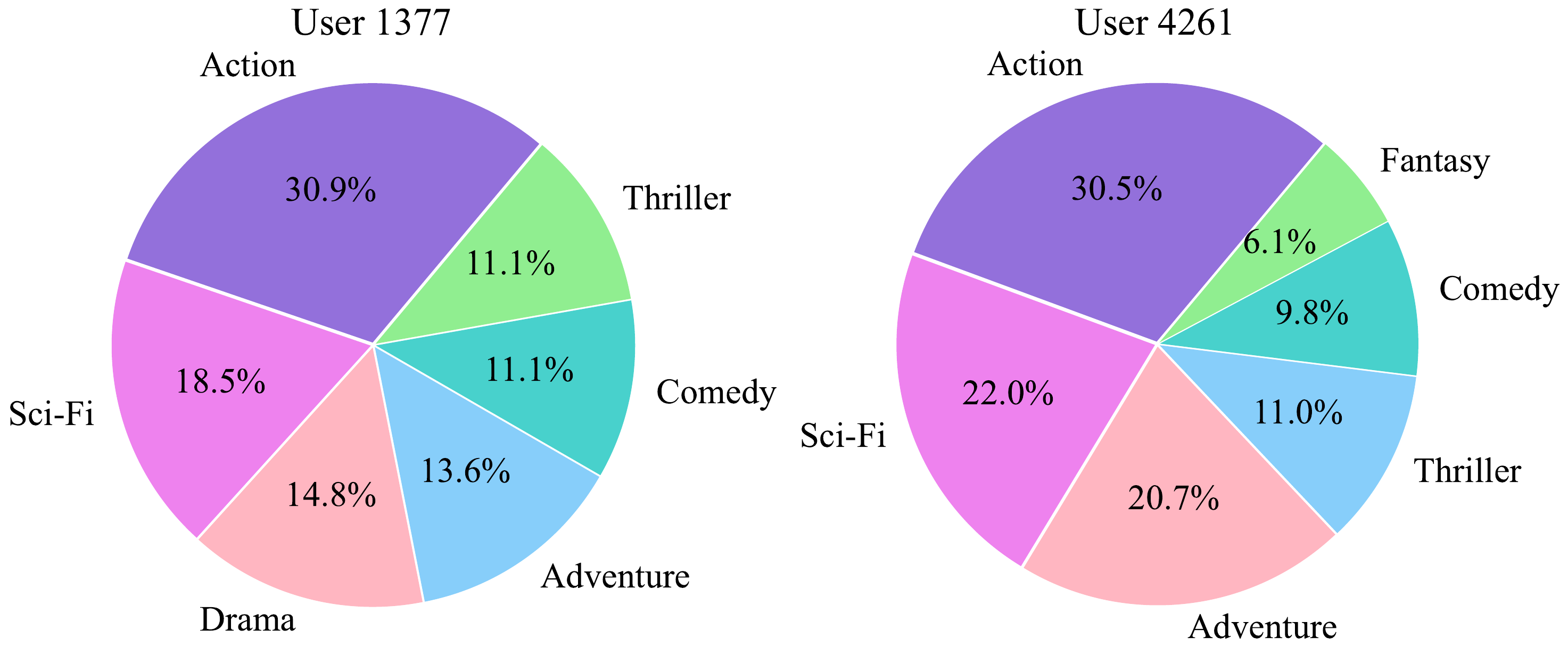}
\caption{The true distributions of the top 6 categories favored by user 1377 and user 4261.}
\label{fig:historical_distribution}
\end{figure}

\begin{table*}[!h]
\begin{minipage}[t]{0.48\linewidth}
\centering
\footnotesize
\centering
\caption{The category of recommended items from LightGCN for user 1377.}
\label{tab:1377_lgcn}
\centering
\begin{tabular}{c|cccc}
\toprule 
\bfseries Item ID& \multicolumn{4}{c}{\bfseries Category}\\
\midrule
248 &\cellcolor[HTML]{F4B9C1}Drama    &War        &&\\
160 &\cellcolor[HTML]{72CECB}Comedy   &\cellcolor[HTML]{F4B9C1}Drama      &&\\
112 &\cellcolor[HTML]{F4B9C1}Drama    &&&\\
364 &\cellcolor[HTML]{8E72D4}Action   &\cellcolor[HTML]{97CCF6}Adventure  &Fantasy     &\cellcolor[HTML]{DF87E8}Sci-Fi\\
252 &\cellcolor[HTML]{F4B9C1}Drama    &\cellcolor[HTML]{A6EC99}Thriller   &&\\
351 &\cellcolor[HTML]{A6EC99}Thriller &&&\\
106 &\cellcolor[HTML]{8E72D4}Action   &\cellcolor[HTML]{97CCF6}Adventure  &&\\
94  &\cellcolor[HTML]{F4B9C1}Drama    &&&\\
1015&Crime    &\cellcolor[HTML]{A6EC99}Thriller   &&\\
72  &\cellcolor[HTML]{8E72D4}Action   &\cellcolor[HTML]{F4B9C1}Drama      &War         &\\
\bottomrule 
\end{tabular}
\end{minipage}
\hfill
\begin{minipage}[t]{0.48\linewidth}
\footnotesize
\centering
\caption{The category of recommended items from \ours-LightGCN for user 1377.}
\label{tab:1377_transfer}
\centering
\begin{tabular}{c|cccc}
\toprule 
\bfseries Item ID& \multicolumn{4}{c}{\bfseries Category}\\
\midrule
160 &\cellcolor[HTML]{72CECB}Comedy   &\cellcolor[HTML]{F4B9C1}Drama      &&\\
364 &\cellcolor[HTML]{8E72D4}Action   &\cellcolor[HTML]{97CCF6}Adventure  &Fantasy     &\cellcolor[HTML]{DF87E8}Sci-Fi\\
372 &Crime    &\cellcolor[HTML]{F4B9C1}Drama      &\cellcolor[HTML]{A6EC99}Thriller    &\\
72  &\cellcolor[HTML]{8E72D4}Action   &\cellcolor[HTML]{F4B9C1}Drama      &War         &\\
252 &\cellcolor[HTML]{F4B9C1}Drama    &\cellcolor[HTML]{A6EC99}Thriller   &&\\
351 &\cellcolor[HTML]{A6EC99}Thriller &&&\\
106 &\cellcolor[HTML]{8E72D4}Action   &\cellcolor[HTML]{97CCF6}Adventure  &&\\
222 &\cellcolor[HTML]{8E72D4}Action   &\cellcolor[HTML]{DF87E8}Sci-Fi     &\cellcolor[HTML]{A6EC99}Thriller    &\\
248 &\cellcolor[HTML]{F4B9C1}Drama    &War        &&\\
82  &\cellcolor[HTML]{8E72D4}Action   &\cellcolor[HTML]{F4B9C1}Drama      &War         &\\
\bottomrule 
\end{tabular}
\label{tab:new_user}
\end{minipage}
\end{table*}

\begin{table*}[!t]
\begin{minipage}[t]{0.455\linewidth}
\centering
\centering
\caption{The category of recommended items from LightGCN for user 4261.}
\label{tab:4261_lgcn}
\centering
\resizebox{\linewidth}{!}{
\begin{tabular}{c|ccccc}
\toprule 
\bfseries Item ID& \multicolumn{5}{c}{\bfseries Category}\\
\midrule
248 &Drama    &War        &&&\\
160 &\cellcolor[HTML]{72CECB}Comedy   &Drama      &&&\\
364 &\cellcolor[HTML]{8E72D4}Action   &\cellcolor[HTML]{F4B9C1}Adventure  &\cellcolor[HTML]{A6EC99}Fantasy     &\cellcolor[HTML]{DF87E8}Sci-Fi&\\
318 &\cellcolor[HTML]{8E72D4}Action   &Crime      &Drama       &&\\
112 &Drama    &&&&\\
252 &Drama    &\cellcolor[HTML]{97CCF6}Thriller   &&&\\
94  &Drama    &&&&\\
1015&Crime    &\cellcolor[HTML]{97CCF6}Thriller   &&&\\
72  &\cellcolor[HTML]{8E72D4}Action   &Drama      &War         &&\\
26  &\cellcolor[HTML]{8E72D4}Action   &\cellcolor[HTML]{F4B9C1}Adventure  &Drama       &\cellcolor[HTML]{DF87E8}Sci-Fi    &War\\
\bottomrule 
\end{tabular}
}
\end{minipage}
\hfill
\begin{minipage}[t]{0.505\linewidth}
\footnotesize
\centering
\caption{The category of recommended items from GraphTransfer-LightGCN for user 4261.}
\label{tab:4261_transfer}
\centering
\resizebox{\linewidth}{!}{
\begin{tabular}{c|ccccc}
\toprule 
\bfseries Item ID& \multicolumn{5}{c}{\bfseries Category}\\
\midrule
160 &\cellcolor[HTML]{72CECB}Comedy   &Drama      &&&\\
364 &\cellcolor[HTML]{8E72D4}Action   &\cellcolor[HTML]{F4B9C1}Adventure  &\cellcolor[HTML]{A6EC99}Fantasy     &\cellcolor[HTML]{DF87E8}Sci-Fi&\\
26  &\cellcolor[HTML]{8E72D4}Action   &\cellcolor[HTML]{F4B9C1}Adventure  &Drama       &\cellcolor[HTML]{DF87E8}Sci-Fi    &War\\
372 &Crime    &Drama      &\cellcolor[HTML]{97CCF6}Thriller    &&\\
72  &\cellcolor[HTML]{8E72D4}Action   &Drama      &War         &&\\
252 &Drama    &\cellcolor[HTML]{97CCF6}Thriller   &&&\\
343 &\cellcolor[HTML]{8E72D4}Action   &\cellcolor[HTML]{F4B9C1}Adventure  &\cellcolor[HTML]{DF87E8}Sci-Fi      &&\\
222 &\cellcolor[HTML]{8E72D4}Action   &\cellcolor[HTML]{DF87E8}Sci-Fi     &\cellcolor[HTML]{97CCF6}Thriller    &&\\
82  &\cellcolor[HTML]{8E72D4}Action   &Drama      &War         &&\\
248 &Drama    &War        &&&\\
\bottomrule 
\end{tabular}
}
\label{tab:new_user}
\end{minipage}
\end{table*}

As seen in Table~\ref{tab:klvalue}, our results show that as $K$ increases from 6 to 14, the KL divergence of \ours-LightGCN is consistently lower than that of LightGCN. This suggests that incorporating auxiliary features improves the alignment of predicted user preferences with their observed historical preferences.

We also assess the impacts of the auxiliary features on recommendation performance by examining the distribution of individual users' preferences. Specifically, Figures~\ref{fig:historical_distribution} shows the distributions of the top 6 categories favored by two randomly chosen users, 1377 and 4261, from the test set. Table~\ref{tab:1377_lgcn} and Table~\ref{tab:1377_transfer} show the categories of top 10 recommended items from LightGCN and \ours-LightGCN for user 1377 respectively, and Table~\ref{tab:4261_lgcn} and Table~\ref{tab:4261_transfer} show the categories of top 10 recommended items from LightGCN and \ours-LightGCN for user 4261 respectively. Comparing the results in Table~\ref{tab:1377_lgcn} and \ref{tab:1377_transfer} with the ground truth in Figure~\ref{fig:historical_distribution}, we find that \ours-LightGCN recommends more Action and Sci-Fi movies that user 1377 is most interested in, and also puts Adventure movies and Comedy movies at higher positions of the recommendation list. Comparing the results in Table~\ref{tab:4261_lgcn} and Table~\ref{tab:4261_transfer} with ground truth in Figure~\ref{fig:historical_distribution}, we can also find that \ours-LightGCN recommends more Action movies, Sci-Fi movies, Adventure movies and Thriller movies that user 4261 is most interested in, and also puts Comedy movies and Fantasy movies at higher positions of the recommendation list. These results further confirm that auxiliary features can effectively improve the accuracy of recommendation results.

In conclusion, from the perspective of both overall distribution and individual user preference, our results show that auxiliary feature extracted by \ours can effectively improve the recommendation performance of existing GNN-based CF methods.

\begin{figure*}[!t]
\centering
\includegraphics[width=5.5in]{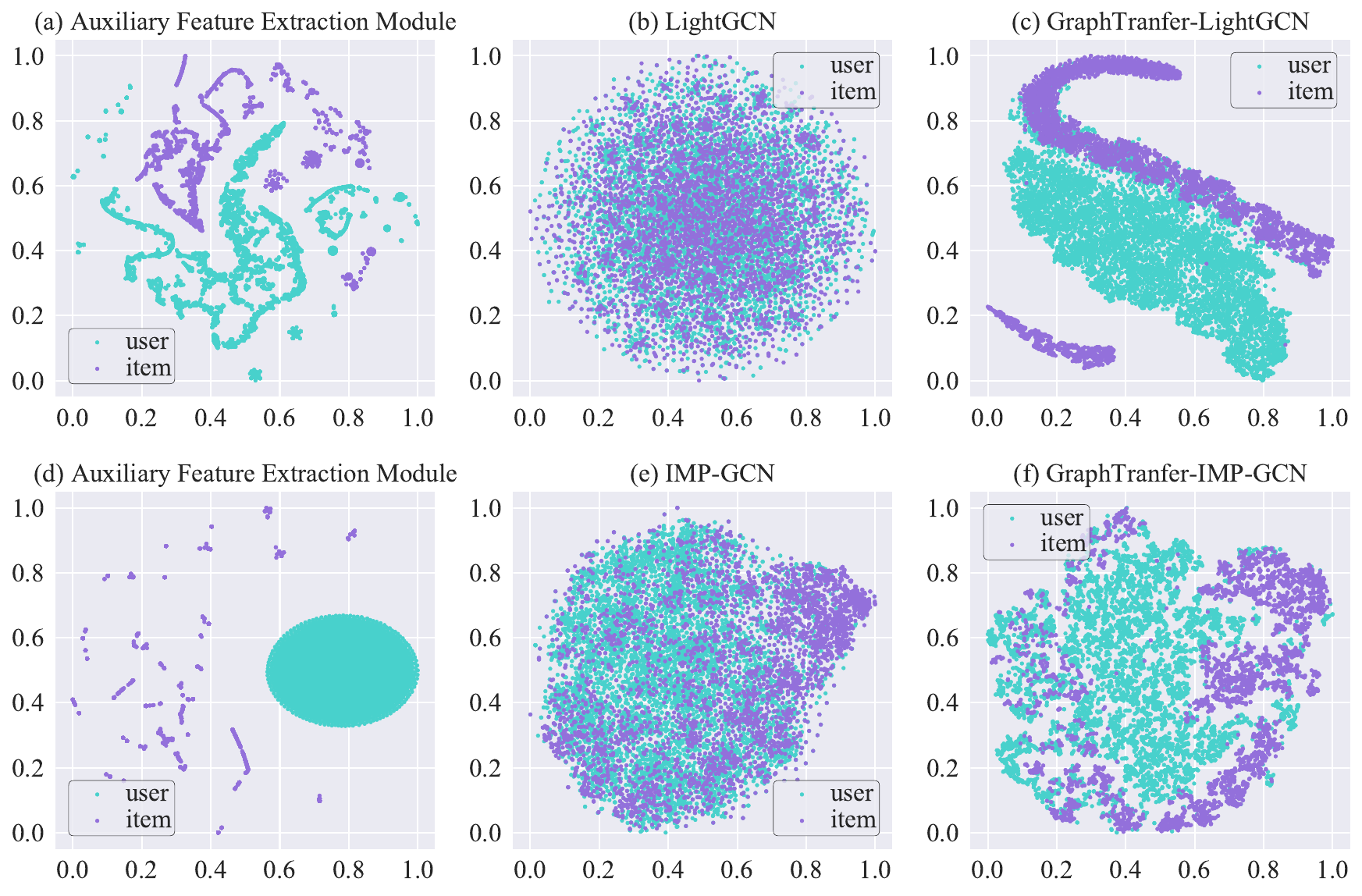}
\caption{The distribution of node embeddings (including users and items) after t-SNE in three different settings. The figure (a) and (d) show the distributions of node embeddings learned by the auxiliary feature extraction module, the figure (b) and (e) show the distributions of node embeddings from the last layer of LightGCN and IMP-GCN, and the figure (c) and (f) show the distributions of node embeddings from the last layer of \ours-LightGCN and \ours-IMP-GCN. 
}
\label{fig:cluster}
\end{figure*}

\subsection{The Over-Smoothing Issue}
One of the major challenges encountered by graph neural networks is the issue of over-smoothing~\cite{oono2019graph}, where node representations tend to converge, making it difficult to differentiate nodes, and in turn leading to reduced performance. In recommender systems, there are two types of node, i.e., user nodes and item nodes, and they are direct neighbors to each other. By iteratively aggregating and updating node representations over the user-item bipartite graph, the representations of users and items will become very similar and lose their unique features, thus obtaining sub-optimal recommendation performance. 
Previous studies have shown that the incorporation of initial residual identity mapping can help alleviate over-smoothing~\cite{gcnii}. We posit that incorporating auxiliary features can also help to alleviate the over-smoothing problem in GNN-based CF algorithms.

We verify our idea with LightGCN and IMP-GCN on ML1M by following~\citet{oversmoothingvisualization} to show how nodes are distributed in the low-dimensional space. Firstly, we obtain the node representations (including users and items) from the last layer of LightGCN and \ours-LightGCN, and IMP-GCN and \ours-IMP-GCN, in which the node representations come from their corresponding graph feature extraction modules. We also obtain the node representations from the auxiliary feature extraction module of \ours-LightGCN and \ours-IMP-GCN. Then, we reduce the dimension of node representations using t-SNE~\cite{tsne}. Figure~\ref{fig:cluster} (a)--(f) show the results in three different settings. 

As seen in Figure~\ref{fig:cluster} (a), the clear separation of users and items is due to their distinct semantic spaces as represented by the auxiliary features. Furthermore, from Figure~\ref{fig:cluster} (b) to (c), the introduction of the auxiliary feature results in a distinct change in the distribution of node representations, making it easier to differentiate users and items. Although iterative aggregation and update will still pull the embeddings of connected user and item pair to closer positions, the cross fusion module will adjust the update directions of user and item embeddings and reduce the loss of their unique features, making them more distinguishable and leading to better performance, as illustrated in Section~\ref{sec:analysis}. Thus, we conclude that, though we cannot address over-smoothing issue completely, using our proposed \ours can help to effectively alleviate over-smoothing issues in graph neural networks. Similar observations can be found from Figure~\ref{fig:cluster} (d)--(f) for IMP-GCN.

\begin{table*}[ht!]
\renewcommand{\arraystretch}{1.3}
\caption{Universality analysis results in three scenarios. The top part shows the results of fusing graph and text features. The middle part shows the results of fusing features from different time periods. Note that JODIE cannot finish training within a reasonable time on ML1M, so that the results are not reported. The bottom part shows the results of fusing features from different models. Note that GIN+LightGCN means the fusion of GIN and LightGCN.}
\label{tab:scal_expr}
\centering
\resizebox{\linewidth}{!}{
\begin{tabular}{l|cccccccccccc}
\toprule
 &\multicolumn{6}{c}{Book} & \multicolumn{6}{c}{Yelp}\\
\cmidrule(r){2-7}\cmidrule(r){8-13}
&F1@5&MRR@5&NDCG@5&F1@10&MRR@10&NDCG@10 &F1@5& MRR@5& NDCG@5&F1@10&MRR@10&NDCG@10\\
\midrule
GIN&0.0113&0.0281&0.0349&0.0127&0.0306&0.0436&0.0069&0.0269&0.0323&0.0107&0.0333&0.0463\\
\ours-GIN&0.0117&0.0281&0.0355&0.0136&0.0311&0.0449&0.0076&0.0293&0.0351&0.0123&0.0369&0.0497\\
Impr.(\%)&+3.54&+0.00&+1.72&+7.09&+1.63&+2.98&+10.14&+8.92&+8.67&+14.95&+10.81&+7.34\\
\midrule
LightGCN&0.0282&0.0557&0.0716&0.0294&0.0542&0.0812&0.0331&0.0927&0.1146&0.0402&0.1032&0.1442\\
\ours-LightGCN&0.0291&0.0575&0.0732&0.0296&0.0556&0.0828&0.0340&0.0948&0.1169&0.0411&0.1046&0.1459\\
Impr.(\%)&+3.09&+3.23&+2.23&+0.68&+2.58&+1.97&+2.72&+2.27&+2.01&+2.24&+1.36&+1.18\\
\midrule
IMP-GCN&0.0301&0.0603&0.0766&0.0319&0.0601&0.0892&0.0270&0.0723&0.0896&0.0308&0.0762&0.1083\\
\ours-IMP-GCN&0.0304&0.0624&0.0789&0.0333&0.0616&0.0917&0.0279&0.0772&0.0949&0.0314&0.0786&0.1115\\
Impr.(\%)&+1.00&+3.48&+3.00&+4.39&+2.50&+2.80&+3.33&+6.78&+5.92&+1.95&+3.15&+2.95\\
\midrule\midrule
 &\multicolumn{6}{c}{Video} & \multicolumn{6}{c}{ML1M}\\
\cmidrule{2-7}\cmidrule{8-13}
&F1@5&MRR@5&NDCG@5&F1@10&MRR@10&NDCG@10 &F1@5& MRR@5& NDCG@5&F1@10&MRR@10&NDCG@10\\
\midrule
RRN&0.0029&0.0032&0.0049&0.0035&0.0047&0.0085&0.0365&0.1155&0.1378&0.0510&0.1325&0.1834\\
\ours-RRN&0.0045&0.0080&0.0106&0.0063&0.0098&0.0162&0.0368&0.1390&0.1603&0.0538&0.1488&0.1959\\
Impr.(\%)&+55.17&+150.00&+116.33&+80.00&+108.51&+90.59&+0.82&+20.34&+16.32&+5.49&+12.30&+6.82\\
\midrule
GCMCRNN&0.0090&0.0076&0.0135&0.0069&0.0059&0.0138&0.0374&0.1625&0.1839&0.0599&0.1682&0.2173\\
\ours-GCMCRNN&0.0098&0.0087&0.0141&0.0074&0.0060&0.0147&0.0452&0.1729&0.2047&0.0651&0.1873&0.2337\\
Impr.(\%)&+8.89&+14.47&+4.44&+7.25&+1.69&+6.52&+20.86&+6.40&+11.31&+8.68&+11.36&+7.55\\
\midrule
JODIE&0.0160&0.0199&0.0294&0.0145&0.0232&0.0374&--&--&--&--&--&--\\
\ours-JODIE&0.0164&0.0228&0.0324&0.0161&0.0273&0.0437&--&--&--&--&--&--\\
Impr.(\%)&+2.50&+14.57&+10.20&+11.03&+17.67&+16.84&--&--&--&--&--&--\\
\midrule\midrule
 &\multicolumn{6}{c}{ML1M} & \multicolumn{6}{c}{Music}\\
\cmidrule(r){2-7}\cmidrule(r){8-13}
&F1@5&MRR@5&NDCG@5&F1@10&MRR@10&NDCG@10 &F1@5& MRR@5& NDCG@5&F1@10&MRR@10&NDCG@10\\
\midrule
GCMC&0.0383&0.1281&0.1603&0.0576&0.1561&0.2124&0.0372&0.3261&0.3680&0.0677&0.3264&0.3972\\
GIN&0.0720&0.2720&0.3153&0.1015&0.2627&0.3364&0.0619&0.5202&0.5744&0.0956&0.4849&0.5594\\
LightGCN&0.0626&0.2872&0.3260&0.0854&0.2912&0.3506&0.0423&0.3782&0.4231&0.0638&0.3625&0.4302\\
IMP-GCN&0.0828&0.2886&0.3375&0.1081&0.2712&0.3483&0.0636&0.5227&0.5789&0.0972&0.4888&0.5661\\
\midrule
GIN+LightGCN&0.0836&0.3052&0.3504&0.1126&0.3018&0.3719&0.0627&0.5441&0.5887&0.0972&0.5117&0.5803\\
Impr.(\%)&+16.11&+6.27&+7.48&+10.94&+3.64&+6.08&+1.29&+4.59&+2.49&+1.67&+5.53&+3.74\\
\midrule
GCMC+LightGCN&0.0759&0.2970&0.3374&0.1052&0.3071&0.3686&0.0624&0.5231&0.5778&0.0961&0.4955&0.5709\\
Impr.(\%)&+21.25&+3.41&+3.50&+23.19&+5.46&+5.13&+47.52&+38.31&+36.56&+41.95&+36.69&+32.71\\
\midrule
LightGCN+IMP-GCN&0.0822&0.3068&0.3509&0.1143&0.3080&0.3780&0.0643&0.5297&0.5850&0.0988&0.5079&0.5797\\
Impr.(\%)&-0.72&+6.31&+3.97&+5.74&+5.77&+7.82&+1.10&+1.34&+1.05&+1.65&+3.91&+2.40\\
\bottomrule
\end{tabular}}
\end{table*}

\section{Universality Experiments}
To further examine the universality of our proposed \ours method, we evaluate its effectiveness in a variety of feature fusion scenarios, including the fusion of graph feature and text feature, the fusion of features in different time periods of sequential models and the fusion of features extracted by different recommendation models.

\subsection{Fusion of Reviews}
In this experiment, we study the performance of \ours in the fusion of graph feature from user-item interactions and text feature from user's reviews. Experiments were conducted on two datasets, Book and Yelp, both containing rich user reviews on items. We selected GIN, LightGCN and IMP-GCN as baselines, as they are limited to extracting only graph features from user interactions. Regarding the text feature, we utilized the NARRE~\cite{narre} method for extraction as it achieves better performance than BERT~\cite{bert} in our experiments.

The experiment results can be seen in the top part of Table~\ref{tab:scal_expr}. We observe that incorporating text feature significantly improves the recommendation accuracy of GIN/LightGCN/IMP-GCN. This is because text feature carries more refined information about users and items such as user age and item popularity inferred from reviews. Thus, it is more beneficial for modeling users and items. Our proposed cross fusion module in \ours can effectively combine graph and text features, resulting in more accurate user and item modeling and better performance in recommendation.

\subsection{Fusion of Sequential Models from Multiple Time Periods}
In this experiment, we investigate the ability of our proposed \ours method to fuse features across different time periods in sequential recommendation models to enhance  accuracy. The experiments were conducted using the Video and ML1M datasets, with RRN, GCMCRNN and JODIE selected as baselines that can capture temporal user and item features from chronological user interactions.

The middle part of Table~\ref{tab:scal_expr} presents the results. We can observe that equipping RRN, GCMCRCN and JODIE with \ours results in improved accuracy for all methods. This is because incorporating features from previous time periods, as discussed in Section~\ref{sec:extension}, which has two advantages: 1) it enhances the model's ability to remember historical user and item features, resulting in a more precise modeling of the user and item; 2) it smooths the changing of the user and item embeddings over time, minimizing the model fluctuations caused by noisy interactions. Our cross fusion module effectively merges features from different time periods, resulting in better recommendation accuracy.

\subsection{Fusion of Features from Multiple Models}
In this experiment, we study whether \ours can fuse features from different models to improve the recommendation accuracy. In this experiment, we choose the ML1M and Music datasets and take GCMC, GIN, LightGCN and IMP-GCN as the baselines to evaluate the performance of \ours.

The bottom part of Table~\ref{tab:scal_expr} shows the results. Compared to utilizing a single model, we found that integrating multiple models, i.e., GIN+LightGCN, GCMC+LightGCN and LightGCN+IMP-GCN, results in better performance. This is because different models capture the characteristics of user and item from different perspectives, hence using them results in more comprehensive modeling of users and items. Our proposed cross fusion module can effectively fuse features from different recommendation models, resulting in more accurate modeling of users and items and thus superior performance when compared to a single model.

\section{Conclusion}
We propose \ours, a universal feature fusion framework for graph neural network based collaborative filtering methods, which seamlessly integrates graph and auxiliary features. We use existing graph embedding techniques to extract graph features from the user-item interaction graph, then employ a graph neural network-based auxiliary feature extraction module to learn auxiliary features from constructed user-user and item-item interaction graphs. Finally, the cross fusion module in \ours employs a ``learn to fuse'' approach to effectively fuse graph and auxiliary features. Theoretical analysis and extensive experiments on public datasets indicate that our method can significantly outperform other fusion techniques in recommendation performance. Additionally, our framework alleviates over-smoothing problems in traditional GNNs by incorporating auxiliary features. Furthermore, we demonstrate the universality of \ours through its application in fusing different features in three other scenarios to significantly improve the performance of collaborative filtering algorithms.

\bibliographystyle{ACM-Reference-Format}
\bibliography{main}


\begin{thebibliography}{76}


\ifx \showCODEN    \undefined \def \showCODEN     #1{\unskip}     \fi
\ifx \showDOI      \undefined \def \showDOI       #1{#1}\fi
\ifx \showISBNx    \undefined \def \showISBNx     #1{\unskip}     \fi
\ifx \showISBNxiii \undefined \def \showISBNxiii  #1{\unskip}     \fi
\ifx \showISSN     \undefined \def \showISSN      #1{\unskip}     \fi
\ifx \showLCCN     \undefined \def \showLCCN      #1{\unskip}     \fi
\ifx \shownote     \undefined \def \shownote      #1{#1}          \fi
\ifx \showarticletitle \undefined \def \showarticletitle #1{#1}   \fi
\ifx \showURL      \undefined \def \showURL       {\relax}        \fi
\providecommand\bibfield[2]{#2}
\providecommand\bibinfo[2]{#2}
\providecommand\natexlab[1]{#1}
\providecommand\showeprint[2][]{arXiv:#2}

\bibitem[B{\k{a}}czkiewicz et~al\mbox{.}(2021)]%
        {rs3}
\bibfield{author}{\bibinfo{person}{Aleksandra B{\k{a}}czkiewicz}, \bibinfo{person}{Bart{\l}omiej Kizielewicz}, \bibinfo{person}{Andrii Shekhovtsov}, \bibinfo{person}{Jaros{\l}aw W{\k{a}}tr{\'o}bski}, {and} \bibinfo{person}{Wojciech Sa{\l}abun}.} \bibinfo{year}{2021}\natexlab{}.
\newblock \showarticletitle{Methodical aspects of MCDM based E-commerce recommender system}.
\newblock \bibinfo{journal}{\emph{Journal of Theoretical and Applied Electronic Commerce Research}} \bibinfo{volume}{16}, \bibinfo{number}{6} (\bibinfo{year}{2021}), \bibinfo{pages}{2192--2229}.
\newblock


\bibitem[Berg et~al\mbox{.}(2018)]%
        {gcmc}
\bibfield{author}{\bibinfo{person}{Rianne van~den Berg}, \bibinfo{person}{Thomas~N Kipf}, {and} \bibinfo{person}{Max Welling}.} \bibinfo{year}{2018}\natexlab{}.
\newblock \showarticletitle{Graph convolutional matrix completion}. In \bibinfo{booktitle}{\emph{Proceedings of the 24th ACM SIGKDD International Conference on Knowledge Discovery and Data Mining}}.
\newblock


\bibitem[Bo et~al\mbox{.}(2021)]%
        {bo2021beyond}
\bibfield{author}{\bibinfo{person}{Deyu Bo}, \bibinfo{person}{Xiao Wang}, \bibinfo{person}{Chuan Shi}, {and} \bibinfo{person}{Huawei Shen}.} \bibinfo{year}{2021}\natexlab{}.
\newblock \showarticletitle{Beyond low-frequency information in graph convolutional networks}. In \bibinfo{booktitle}{\emph{Proceedings of the AAAI Conference on Artificial Intelligence}}, Vol.~\bibinfo{volume}{35}. \bibinfo{pages}{3950--3957}.
\newblock


\bibitem[Chang et~al\mbox{.}(2020)]%
        {poi1}
\bibfield{author}{\bibinfo{person}{Buru Chang}, \bibinfo{person}{Gwanghoon Jang}, \bibinfo{person}{Seoyoon Kim}, {and} \bibinfo{person}{Jaewoo Kang}.} \bibinfo{year}{2020}\natexlab{}.
\newblock \showarticletitle{Learning Graph-Based Geographical Latent Representation for Point-of-Interest Recommendation}. In \bibinfo{booktitle}{\emph{Proceedings of the 29th ACM International Conference on Information and Knowledge Management}}. \bibinfo{pages}{135–144}.
\newblock


\bibitem[Chang et~al\mbox{.}(2021)]%
        {surge}
\bibfield{author}{\bibinfo{person}{Jianxin Chang}, \bibinfo{person}{Chen Gao}, \bibinfo{person}{Yu Zheng}, \bibinfo{person}{Yiqun Hui}, \bibinfo{person}{Yanan Niu}, \bibinfo{person}{Yang Song}, \bibinfo{person}{Depeng Jin}, {and} \bibinfo{person}{Yong Li}.} \bibinfo{year}{2021}\natexlab{}.
\newblock \showarticletitle{Sequential recommendation with graph neural networks}. In \bibinfo{booktitle}{\emph{Proceedings of the 44th international ACM SIGIR conference on research and development in information retrieval}}. \bibinfo{pages}{378--387}.
\newblock


\bibitem[Chen et~al\mbox{.}(2021)]%
        {ghcf}
\bibfield{author}{\bibinfo{person}{Chong Chen}, \bibinfo{person}{Weizhi Ma}, \bibinfo{person}{Min Zhang}, \bibinfo{person}{Zhaowei Wang}, \bibinfo{person}{Xiuqiang He}, \bibinfo{person}{Chenyang Wang}, \bibinfo{person}{Yiqun Liu}, {and} \bibinfo{person}{Shaoping Ma}.} \bibinfo{year}{2021}\natexlab{}.
\newblock \showarticletitle{Graph heterogeneous multi-relational recommendation}. In \bibinfo{booktitle}{\emph{Proceedings of the AAAI Conference on Artificial Intelligence}}, Vol.~\bibinfo{volume}{35}. \bibinfo{pages}{3958--3966}.
\newblock


\bibitem[Chen et~al\mbox{.}(2018)]%
        {narre}
\bibfield{author}{\bibinfo{person}{Chong Chen}, \bibinfo{person}{Min Zhang}, \bibinfo{person}{Yiqun Liu}, {and} \bibinfo{person}{Shaoping Ma}.} \bibinfo{year}{2018}\natexlab{}.
\newblock \showarticletitle{Neural attentional rating regression with review-level explanations}. In \bibinfo{booktitle}{\emph{Proceedings of the 2018 world wide web conference}}. \bibinfo{pages}{1583--1592}.
\newblock


\bibitem[Chen et~al\mbox{.}(2020b)]%
        {lrgccf}
\bibfield{author}{\bibinfo{person}{Lei Chen}, \bibinfo{person}{Le Wu}, \bibinfo{person}{Richang Hong}, \bibinfo{person}{Kun Zhang}, {and} \bibinfo{person}{Meng Wang}.} \bibinfo{year}{2020}\natexlab{b}.
\newblock \showarticletitle{Revisiting graph based collaborative filtering: A linear residual graph convolutional network approach}. In \bibinfo{booktitle}{\emph{Proceedings of the AAAI conference on artificial intelligence}}, Vol.~\bibinfo{volume}{34}. \bibinfo{pages}{27--34}.
\newblock


\bibitem[Chen et~al\mbox{.}(2020a)]%
        {gcnii}
\bibfield{author}{\bibinfo{person}{Ming Chen}, \bibinfo{person}{Zhewei Wei}, \bibinfo{person}{Zengfeng Huang}, \bibinfo{person}{Bolin Ding}, {and} \bibinfo{person}{Yaliang Li}.} \bibinfo{year}{2020}\natexlab{a}.
\newblock \showarticletitle{Simple and deep graph convolutional networks}. In \bibinfo{booktitle}{\emph{International Conference on Machine Learning}}. \bibinfo{pages}{1725--1735}.
\newblock


\bibitem[Dai et~al\mbox{.}(2021b)]%
        {attention}
\bibfield{author}{\bibinfo{person}{Feifei Dai}, \bibinfo{person}{Xiaoyan Gu}, \bibinfo{person}{Zhuo Wang}, \bibinfo{person}{Bo Li}, \bibinfo{person}{Mingda Qian}, {and} \bibinfo{person}{Weiping Wang}.} \bibinfo{year}{2021}\natexlab{b}.
\newblock \showarticletitle{Attention-Based Multi-view Feature Fusion for Cross-Domain Recommendation}. In \bibinfo{booktitle}{\emph{International Conference on Artificial Neural Networks}}. \bibinfo{pages}{204–216}.
\newblock


\bibitem[Dai et~al\mbox{.}(2016)]%
        {deepcoevolve}
\bibfield{author}{\bibinfo{person}{Hanjun Dai}, \bibinfo{person}{Yichen Wang}, \bibinfo{person}{Rakshit Trivedi}, {and} \bibinfo{person}{Le Song}.} \bibinfo{year}{2016}\natexlab{}.
\newblock \showarticletitle{Deep coevolutionary network: Embedding user and item features for recommendation}.
\newblock \bibinfo{journal}{\emph{arXiv preprint arXiv:1609.03675}} (\bibinfo{year}{2016}).
\newblock


\bibitem[Dai et~al\mbox{.}(2021a)]%
        {affniaff}
\bibfield{author}{\bibinfo{person}{Yimian Dai}, \bibinfo{person}{Fabian Gieseke}, \bibinfo{person}{Stefan Oehmcke}, \bibinfo{person}{Yiquan Wu}, {and} \bibinfo{person}{Kobus Barnard}.} \bibinfo{year}{2021}\natexlab{a}.
\newblock \showarticletitle{Attentional feature fusion}. In \bibinfo{booktitle}{\emph{Proceedings of the IEEE/CVF Winter Conference on Applications of Computer Vision}}. \bibinfo{pages}{3560--3569}.
\newblock


\bibitem[Deldjoo et~al\mbox{.}(2016)]%
        {summationfusion}
\bibfield{author}{\bibinfo{person}{Yashar Deldjoo}, \bibinfo{person}{Mehdi Elahi}, \bibinfo{person}{Paolo Cremonesi}, \bibinfo{person}{Farshad~Bakhshandegan Moghaddam}, {and} \bibinfo{person}{Andrea Luigi~Edoardo Caielli}.} \bibinfo{year}{2016}\natexlab{}.
\newblock \showarticletitle{How to combine visual features with tags to improve movie recommendation accuracy?}. In \bibinfo{booktitle}{\emph{International conference on electronic commerce and web technologies}}. \bibinfo{pages}{34--45}.
\newblock


\bibitem[Dwivedi and Roshni(2017)]%
        {rs4}
\bibfield{author}{\bibinfo{person}{Surabhi Dwivedi} {and} \bibinfo{person}{VS~Kumari Roshni}.} \bibinfo{year}{2017}\natexlab{}.
\newblock \showarticletitle{Recommender system for big data in education}. In \bibinfo{booktitle}{\emph{2017 5th National Conference on E-Learning and E-Learning Technologies (ELELTECH)}}. IEEE, \bibinfo{pages}{1--4}.
\newblock


\bibitem[Fadel and Torres(2018)]%
        {gcmcrnn}
\bibfield{author}{\bibinfo{person}{Samuel~G Fadel} {and} \bibinfo{person}{Ricardo da~S Torres}.} \bibinfo{year}{2018}\natexlab{}.
\newblock \showarticletitle{Link Prediction in Dynamic Graphs for Recommendation}.
\newblock \bibinfo{journal}{\emph{arXiv preprint arXiv:1811.07174}} (\bibinfo{year}{2018}).
\newblock


\bibitem[Fan et~al\mbox{.}(2019)]%
        {graphrec}
\bibfield{author}{\bibinfo{person}{Wenqi Fan}, \bibinfo{person}{Yao Ma}, \bibinfo{person}{Qing Li}, \bibinfo{person}{Yuan He}, \bibinfo{person}{Eric Zhao}, \bibinfo{person}{Jiliang Tang}, {and} \bibinfo{person}{Dawei Yin}.} \bibinfo{year}{2019}\natexlab{}.
\newblock \showarticletitle{Graph Neural Networks for Social Recommendation}. In \bibinfo{booktitle}{\emph{The World Wide Web Conference}}. \bibinfo{pages}{417–426}.
\newblock


\bibitem[Feng et~al\mbox{.}(2022a)]%
        {spacefusion1}
\bibfield{author}{\bibinfo{person}{Panpan Feng}, \bibinfo{person}{Jie Fu}, \bibinfo{person}{Zhaoyang Ge}, \bibinfo{person}{Haiyan Wang}, \bibinfo{person}{Yanjie Zhou}, \bibinfo{person}{Bing Zhou}, {and} \bibinfo{person}{Zongmin Wang}.} \bibinfo{year}{2022}\natexlab{a}.
\newblock \showarticletitle{Unsupervised semantic-aware adaptive feature fusion network for arrhythmia detection}.
\newblock \bibinfo{journal}{\emph{Information Sciences}}  \bibinfo{volume}{582} (\bibinfo{year}{2022}).
\newblock


\bibitem[Feng et~al\mbox{.}(2022b)]%
        {grato}
\bibfield{author}{\bibinfo{person}{Xinshun Feng}, \bibinfo{person}{Herun Wan}, \bibinfo{person}{Shangbin Feng}, \bibinfo{person}{Hongrui Wang}, \bibinfo{person}{Qinghua Zheng}, \bibinfo{person}{Jun Zhou}, {and} \bibinfo{person}{Minnan Luo}.} \bibinfo{year}{2022}\natexlab{b}.
\newblock \showarticletitle{GraTO: Graph Neural Network Framework Tackling Over-smoothing with Neural Architecture Search}. In \bibinfo{booktitle}{\emph{Proceedings of the 31st ACM International Conference on Information ans Knowledge Management}}. \bibinfo{pages}{520--529}.
\newblock


\bibitem[Fu et~al\mbox{.}(2020)]%
        {fu2020fairness}
\bibfield{author}{\bibinfo{person}{Zuohui Fu}, \bibinfo{person}{Yikun Xian}, \bibinfo{person}{Ruoyuan Gao}, \bibinfo{person}{Jieyu Zhao}, \bibinfo{person}{Qiaoying Huang}, \bibinfo{person}{Yingqiang Ge}, \bibinfo{person}{Shuyuan Xu}, \bibinfo{person}{Shijie Geng}, \bibinfo{person}{Chirag Shah}, \bibinfo{person}{Yongfeng Zhang}, {et~al\mbox{.}}} \bibinfo{year}{2020}\natexlab{}.
\newblock \showarticletitle{Fairness-aware explainable recommendation over knowledge graphs}. In \bibinfo{booktitle}{\emph{Proceedings of the 43rd International ACM SIGIR Conference on Research and Development in Information Retrieval}}. \bibinfo{pages}{69--78}.
\newblock


\bibitem[Gao et~al\mbox{.}(2022)]%
        {social3}
\bibfield{author}{\bibinfo{person}{Chen Gao}, \bibinfo{person}{Xiang Wang}, \bibinfo{person}{Xiangnan He}, {and} \bibinfo{person}{Yong Li}.} \bibinfo{year}{2022}\natexlab{}.
\newblock \showarticletitle{Graph neural networks for recommender system}. In \bibinfo{booktitle}{\emph{Proceedings of the Fifteenth ACM International Conference on Web Search and Data Mining}}. \bibinfo{pages}{1623--1625}.
\newblock


\bibitem[Gao et~al\mbox{.}(2019)]%
        {spacefusion2}
\bibfield{author}{\bibinfo{person}{Xizhan Gao}, \bibinfo{person}{Quansen Sun}, \bibinfo{person}{Haitao Xu}, \bibinfo{person}{Dong Wei}, {and} \bibinfo{person}{Jianqiang Gao}.} \bibinfo{year}{2019}\natexlab{}.
\newblock \showarticletitle{Multi-model fusion metric learning for image set classification}.
\newblock \bibinfo{journal}{\emph{Knowledge-Based Systems}}  \bibinfo{volume}{164} (\bibinfo{year}{2019}), \bibinfo{pages}{253--264}.
\newblock


\bibitem[Guo et~al\mbox{.}(2023)]%
        {seqrecref1}
\bibfield{author}{\bibinfo{person}{Naicheng Guo}, \bibinfo{person}{Xiaolei Liu}, \bibinfo{person}{Shaoshuai Li}, \bibinfo{person}{Qiongxu Ma}, \bibinfo{person}{Kaixin Gao}, \bibinfo{person}{Bing Han}, \bibinfo{person}{Lin Zheng}, \bibinfo{person}{Sheng Guo}, {and} \bibinfo{person}{Xiaobo Guo}.} \bibinfo{year}{2023}\natexlab{}.
\newblock \showarticletitle{Poincar{\'e} Heterogeneous Graph Neural Networks for Sequential Recommendation}.
\newblock \bibinfo{journal}{\emph{ACM Transactions on Information Systems}} \bibinfo{volume}{41}, \bibinfo{number}{3} (\bibinfo{year}{2023}), \bibinfo{pages}{1--26}.
\newblock


\bibitem[Gupta et~al\mbox{.}(2020)]%
        {rs2}
\bibfield{author}{\bibinfo{person}{Meenu Gupta}, \bibinfo{person}{Aditya Thakkar}, \bibinfo{person}{Vishal Gupta}, \bibinfo{person}{Dhruv Pratap~Singh Rathore}, {et~al\mbox{.}}} \bibinfo{year}{2020}\natexlab{}.
\newblock \showarticletitle{Movie recommender system using collaborative filtering}. In \bibinfo{booktitle}{\emph{2020 International Conference on Electronics and Sustainable Communication Systems (ICESC)}}. IEEE, \bibinfo{pages}{415--420}.
\newblock


\bibitem[He et~al\mbox{.}(2020)]%
        {lightgcn}
\bibfield{author}{\bibinfo{person}{Xiangnan He}, \bibinfo{person}{Kuan Deng}, \bibinfo{person}{Xiang Wang}, \bibinfo{person}{Yan Li}, \bibinfo{person}{YongDong Zhang}, {and} \bibinfo{person}{Meng Wang}.} \bibinfo{year}{2020}\natexlab{}.
\newblock \showarticletitle{LightGCN: Simplifying and Powering Graph Convolution Network for Recommendation}. In \bibinfo{booktitle}{\emph{Proceedings of the 43rd International ACM SIGIR Conference on Research and Development in Information Retrieval}}. \bibinfo{pages}{639–648}.
\newblock


\bibitem[Hsu and Li(2021)]%
        {retagnn}
\bibfield{author}{\bibinfo{person}{Cheng Hsu} {and} \bibinfo{person}{Cheng-Te Li}.} \bibinfo{year}{2021}\natexlab{}.
\newblock \showarticletitle{RetaGNN: relational temporal attentive graph neural networks for holistic sequential recommendation}. In \bibinfo{booktitle}{\emph{Proceedings of the web conference 2021}}. \bibinfo{pages}{2968--2979}.
\newblock


\bibitem[Ioffe and Szegedy(2015)]%
        {batchnorm}
\bibfield{author}{\bibinfo{person}{Sergey Ioffe} {and} \bibinfo{person}{Christian Szegedy}.} \bibinfo{year}{2015}\natexlab{}.
\newblock \showarticletitle{Batch normalization: Accelerating deep network training by reducing internal covariate shift}. In \bibinfo{booktitle}{\emph{International conference on machine learning}}. pmlr, \bibinfo{pages}{448--456}.
\newblock


\bibitem[Kenton and Toutanova(2019)]%
        {bert}
\bibfield{author}{\bibinfo{person}{Jacob Devlin Ming-Wei~Chang Kenton} {and} \bibinfo{person}{Lee~Kristina Toutanova}.} \bibinfo{year}{2019}\natexlab{}.
\newblock \showarticletitle{BERT: Pre-training of Deep Bidirectional Transformers for Language Understanding}. In \bibinfo{booktitle}{\emph{Proceedings of NAACL-HLT}}. \bibinfo{pages}{4171--4186}.
\newblock


\bibitem[Krizhevsky et~al\mbox{.}(2012)]%
        {krizhevsky2012imagenet}
\bibfield{author}{\bibinfo{person}{Alex Krizhevsky}, \bibinfo{person}{Ilya Sutskever}, {and} \bibinfo{person}{Geoffrey~E Hinton}.} \bibinfo{year}{2012}\natexlab{}.
\newblock \showarticletitle{Imagenet classification with deep convolutional neural networks}.
\newblock \bibinfo{journal}{\emph{Advances in neural information processing systems}}  \bibinfo{volume}{25} (\bibinfo{year}{2012}).
\newblock


\bibitem[Kumar et~al\mbox{.}(2019)]%
        {jodie}
\bibfield{author}{\bibinfo{person}{Srijan Kumar}, \bibinfo{person}{Xikun Zhang}, {and} \bibinfo{person}{Jure Leskovec}.} \bibinfo{year}{2019}\natexlab{}.
\newblock \showarticletitle{Predicting Dynamic Embedding Trajectory in Temporal Interaction Networks}. In \bibinfo{booktitle}{\emph{Proceedings of the 25th ACM SIGKDD International Conference on Knowledge Discovery and Data Mining}}. \bibinfo{pages}{1269–1278}.
\newblock


\bibitem[Li et~al\mbox{.}(2020)]%
        {li2020learning}
\bibfield{author}{\bibinfo{person}{Xiangsheng Li}, \bibinfo{person}{Maarten de Rijke}, \bibinfo{person}{Yiqun Liu}, \bibinfo{person}{Jiaxin Mao}, \bibinfo{person}{Weizhi Ma}, \bibinfo{person}{Min Zhang}, {and} \bibinfo{person}{Shaoping Ma}.} \bibinfo{year}{2020}\natexlab{}.
\newblock \showarticletitle{Learning better representations for neural information retrieval with graph information}. In \bibinfo{booktitle}{\emph{Proceedings of the 29th ACM International Conference on Information and Knowledge Management}}. \bibinfo{pages}{795--804}.
\newblock


\bibitem[Liang et~al\mbox{.}(2021)]%
        {survey2}
\bibfield{author}{\bibinfo{person}{Zijian Liang}, \bibinfo{person}{Hui Ding}, {and} \bibinfo{person}{Wenlong Fu}.} \bibinfo{year}{2021}\natexlab{}.
\newblock \showarticletitle{A Survey on Graph Neural Networks for Recommendation}. In \bibinfo{booktitle}{\emph{2021 International Conference on Culture-oriented Science Technology (ICCST)}}. \bibinfo{pages}{383--386}.
\newblock


\bibitem[Lim et~al\mbox{.}(2022)]%
        {poi3}
\bibfield{author}{\bibinfo{person}{Nicholas Lim}, \bibinfo{person}{Bryan Hooi}, \bibinfo{person}{See-Kiong Ng}, \bibinfo{person}{Yong~Liang Goh}, \bibinfo{person}{Renrong Weng}, {and} \bibinfo{person}{Rui Tan}.} \bibinfo{year}{2022}\natexlab{}.
\newblock \showarticletitle{Hierarchical multi-task graph recurrent network for next poi recommendation}. In \bibinfo{booktitle}{\emph{Proceedings of the 45th international ACM SIGIR conference on Research and development in Information Retrieval}}.
\newblock


\bibitem[Lim et~al\mbox{.}(2020a)]%
        {poi2}
\bibfield{author}{\bibinfo{person}{Nicholas Lim}, \bibinfo{person}{Bryan Hooi}, \bibinfo{person}{See-Kiong Ng}, \bibinfo{person}{Xueou Wang}, \bibinfo{person}{Yong~Liang Goh}, \bibinfo{person}{Renrong Weng}, {and} \bibinfo{person}{Jagannadan Varadarajan}.} \bibinfo{year}{2020}\natexlab{a}.
\newblock \showarticletitle{STP-UDGAT: Spatial-Temporal-Preference User Dimensional Graph Attention Network for Next POI Recommendation}. In \bibinfo{booktitle}{\emph{Proceedings of the 29th ACM International Conference on Information and Knowledge Management}}. \bibinfo{pages}{845–854}.
\newblock


\bibitem[Lim et~al\mbox{.}(2020b)]%
        {stpudgat}
\bibfield{author}{\bibinfo{person}{Nicholas Lim}, \bibinfo{person}{Bryan Hooi}, \bibinfo{person}{See-Kiong Ng}, \bibinfo{person}{Xueou Wang}, \bibinfo{person}{Yong~Liang Goh}, \bibinfo{person}{Renrong Weng}, {and} \bibinfo{person}{Jagannadan Varadarajan}.} \bibinfo{year}{2020}\natexlab{b}.
\newblock \showarticletitle{STP-UDGAT: spatial-temporal-preference user dimensional graph attention network for next POI recommendation}. In \bibinfo{booktitle}{\emph{Proceedings of the 29th ACM International Conference on Information and Knowledge Management}}. \bibinfo{pages}{845--854}.
\newblock


\bibitem[Liu et~al\mbox{.}(2021)]%
        {impgcn}
\bibfield{author}{\bibinfo{person}{Fan Liu}, \bibinfo{person}{Zhiyong Cheng}, \bibinfo{person}{Lei Zhu}, \bibinfo{person}{Zan Gao}, {and} \bibinfo{person}{Liqiang Nie}.} \bibinfo{year}{2021}\natexlab{}.
\newblock \showarticletitle{Interest-Aware Message-Passing GCN for Recommendation}. In \bibinfo{booktitle}{\emph{Proceedings of the Web Conference 2021}}. \bibinfo{pages}{1296–1305}.
\newblock


\bibitem[Liu et~al\mbox{.}(2020b)]%
        {liu2020inter}
\bibfield{author}{\bibinfo{person}{Feng Liu}, \bibinfo{person}{Weiwen Liu}, \bibinfo{person}{Xutao Li}, {and} \bibinfo{person}{Yunming Ye}.} \bibinfo{year}{2020}\natexlab{b}.
\newblock \showarticletitle{Inter-sequence Enhanced Framework for Personalized Sequential Recommendation}. In \bibinfo{booktitle}{\emph{Proceedings of the AAAI Conference on Artificial Intelligence}}.
\newblock


\bibitem[Liu et~al\mbox{.}(2022)]%
        {gnncfref2}
\bibfield{author}{\bibinfo{person}{Kang Liu}, \bibinfo{person}{Feng Xue}, \bibinfo{person}{Dan Guo}, \bibinfo{person}{Le Wu}, \bibinfo{person}{Shujie Li}, {and} \bibinfo{person}{Richang Hong}.} \bibinfo{year}{2022}\natexlab{}.
\newblock \showarticletitle{MEGCF: Multimodal entity graph collaborative filtering for personalized recommendation}.
\newblock \bibinfo{journal}{\emph{ACM Transactions on Information Systems (TOIS)}} (\bibinfo{year}{2022}).
\newblock


\bibitem[Liu et~al\mbox{.}(2020a)]%
        {liu2020towards}
\bibfield{author}{\bibinfo{person}{Meng Liu}, \bibinfo{person}{Hongyang Gao}, {and} \bibinfo{person}{Shuiwang Ji}.} \bibinfo{year}{2020}\natexlab{a}.
\newblock \showarticletitle{Towards deeper graph neural networks}. In \bibinfo{booktitle}{\emph{Proceedings of the 26th ACM SIGKDD international conference on knowledge discovery and data mining}}. \bibinfo{pages}{338--348}.
\newblock


\bibitem[Lyu et~al\mbox{.}(2022a)]%
        {interpret3}
\bibfield{author}{\bibinfo{person}{Ziyu Lyu}, \bibinfo{person}{Yue Wu}, \bibinfo{person}{Junjie Lai}, \bibinfo{person}{Min Yang}, \bibinfo{person}{Chengming Li}, {and} \bibinfo{person}{Wei Zhou}.} \bibinfo{year}{2022}\natexlab{a}.
\newblock \showarticletitle{Knowledge Enhanced Graph Neural Networks for Explainable Recommendation}.
\newblock \bibinfo{journal}{\emph{IEEE Transactions on Knowledge and Data Engineering}} (\bibinfo{year}{2022}).
\newblock


\bibitem[Lyu et~al\mbox{.}(2022b)]%
        {lyu2022knowledge}
\bibfield{author}{\bibinfo{person}{Ziyu Lyu}, \bibinfo{person}{Yue Wu}, \bibinfo{person}{Junjie Lai}, \bibinfo{person}{Min Yang}, \bibinfo{person}{Chengming Li}, {and} \bibinfo{person}{Wei Zhou}.} \bibinfo{year}{2022}\natexlab{b}.
\newblock \showarticletitle{Knowledge Enhanced Graph Neural Networks for Explainable Recommendation}.
\newblock \bibinfo{journal}{\emph{IEEE Transactions on Knowledge and Data Engineering}} (\bibinfo{year}{2022}).
\newblock


\bibitem[Niu et~al\mbox{.}(2021)]%
        {niu2021review}
\bibfield{author}{\bibinfo{person}{Zhaoyang Niu}, \bibinfo{person}{Guoqiang Zhong}, {and} \bibinfo{person}{Hui Yu}.} \bibinfo{year}{2021}\natexlab{}.
\newblock \showarticletitle{A review on the attention mechanism of deep learning}.
\newblock \bibinfo{journal}{\emph{Neurocomputing}}  \bibinfo{volume}{452} (\bibinfo{year}{2021}), \bibinfo{pages}{48--62}.
\newblock


\bibitem[Obeid et~al\mbox{.}(2018)]%
        {rs5}
\bibfield{author}{\bibinfo{person}{Charbel Obeid}, \bibinfo{person}{Inaya Lahoud}, \bibinfo{person}{Hicham El~Khoury}, {and} \bibinfo{person}{Pierre-Antoine Champin}.} \bibinfo{year}{2018}\natexlab{}.
\newblock \showarticletitle{Ontology-based recommender system in higher education}. In \bibinfo{booktitle}{\emph{Companion proceedings of the the web conference 2018}}. \bibinfo{pages}{1031--1034}.
\newblock


\bibitem[Oono and Suzuki(2019)]%
        {oono2019graph}
\bibfield{author}{\bibinfo{person}{Kenta Oono} {and} \bibinfo{person}{Taiji Suzuki}.} \bibinfo{year}{2019}\natexlab{}.
\newblock \showarticletitle{Graph Neural Networks Exponentially Lose Expressive Power for Node Classification}. In \bibinfo{booktitle}{\emph{International Conference on Learning Representations}}.
\newblock


\bibitem[Peng et~al\mbox{.}(2022)]%
        {svdgcn}
\bibfield{author}{\bibinfo{person}{Shaowen Peng}, \bibinfo{person}{Kazunari Sugiyama}, {and} \bibinfo{person}{Tsunenori Mine}.} \bibinfo{year}{2022}\natexlab{}.
\newblock \showarticletitle{SVD-GCN: A Simplified Graph Convolution Paradigm for Recommendation}. In \bibinfo{booktitle}{\emph{Proceedings of the 31st ACM International Conference on Information and Knowledge Management}}. \bibinfo{pages}{1625--1634}.
\newblock


\bibitem[Qin et~al\mbox{.}(2020)]%
        {ffanet}
\bibfield{author}{\bibinfo{person}{Xu Qin}, \bibinfo{person}{Zhilin Wang}, \bibinfo{person}{Yuanchao Bai}, \bibinfo{person}{Xiaodong Xie}, {and} \bibinfo{person}{Huizhu Jia}.} \bibinfo{year}{2020}\natexlab{}.
\newblock \showarticletitle{FFA-Net: Feature fusion attention network for single image dehazing}. In \bibinfo{booktitle}{\emph{Proceedings of the AAAI conference on artificial intelligence}}, Vol.~\bibinfo{volume}{34}. \bibinfo{pages}{11908--11915}.
\newblock


\bibitem[Quijano-S{\'a}nchez et~al\mbox{.}(2020)]%
        {rs1}
\bibfield{author}{\bibinfo{person}{Lara Quijano-S{\'a}nchez}, \bibinfo{person}{Iv{\'a}n Cantador}, \bibinfo{person}{Mar{\'\i}a~E Cort{\'e}s-Cediel}, {and} \bibinfo{person}{Olga Gil}.} \bibinfo{year}{2020}\natexlab{}.
\newblock \showarticletitle{Recommender systems for smart cities}.
\newblock \bibinfo{journal}{\emph{Information systems}}  \bibinfo{volume}{92} (\bibinfo{year}{2020}), \bibinfo{pages}{101545}.
\newblock


\bibitem[Rendle et~al\mbox{.}(2009)]%
        {bpr}
\bibfield{author}{\bibinfo{person}{Steffen Rendle}, \bibinfo{person}{Christoph Freudenthaler}, \bibinfo{person}{Zeno Gantner}, {and} \bibinfo{person}{Lars Schmidt-Thieme}.} \bibinfo{year}{2009}\natexlab{}.
\newblock \showarticletitle{BPR: Bayesian Personalized Ranking from Implicit Feedback}. In \bibinfo{booktitle}{\emph{Proceedings of the Twenty-Fifth Conference on Uncertainty in Artificial Intelligence}}. \bibinfo{pages}{452–461}.
\newblock


\bibitem[Roy and Ding(2021)]%
        {roy2021multi}
\bibfield{author}{\bibinfo{person}{Debashish Roy} {and} \bibinfo{person}{Chen Ding}.} \bibinfo{year}{2021}\natexlab{}.
\newblock \showarticletitle{Multi-source based movie recommendation with ratings and the side information}.
\newblock \bibinfo{journal}{\emph{Social Network Analysis and Mining}} \bibinfo{volume}{11}, \bibinfo{number}{1} (\bibinfo{year}{2021}), \bibinfo{pages}{1--20}.
\newblock


\bibitem[Shen et~al\mbox{.}(2021)]%
        {gfcf}
\bibfield{author}{\bibinfo{person}{Yifei Shen}, \bibinfo{person}{Yongji Wu}, \bibinfo{person}{Yao Zhang}, \bibinfo{person}{Caihua Shan}, \bibinfo{person}{Jun Zhang}, \bibinfo{person}{B.~Khaled Letaief}, {and} \bibinfo{person}{Dongsheng Li}.} \bibinfo{year}{2021}\natexlab{}.
\newblock \showarticletitle{How Powerful is Graph Convolution for Recommendation?}. In \bibinfo{booktitle}{\emph{Proceedings of the 30th ACM International Conference on Information and Knowledge Management}}. \bibinfo{pages}{1619–1629}.
\newblock


\bibitem[Shi et~al\mbox{.}(2019)]%
        {hin}
\bibfield{author}{\bibinfo{person}{Chuan Shi}, \bibinfo{person}{Binbin Hu}, \bibinfo{person}{Wayne~Xin Zhao}, {and} \bibinfo{person}{Philip~S. Yu}.} \bibinfo{year}{2019}\natexlab{}.
\newblock \showarticletitle{Heterogeneous Information Network Embedding for Recommendation}.
\newblock \bibinfo{journal}{\emph{IEEE Transactions on Knowledge and Data Engineering}} \bibinfo{volume}{31}, \bibinfo{number}{2} (\bibinfo{year}{2019}), \bibinfo{pages}{357--370}.
\newblock


\bibitem[Shi et~al\mbox{.}(2017)]%
        {hin_survey}
\bibfield{author}{\bibinfo{person}{Chuan Shi}, \bibinfo{person}{Yitong Li}, \bibinfo{person}{Jiawei Zhang}, \bibinfo{person}{Yizhou Sun}, {and} \bibinfo{person}{Philip~S. Yu}.} \bibinfo{year}{2017}\natexlab{}.
\newblock \showarticletitle{A survey of heterogeneous information network analysis}.
\newblock \bibinfo{journal}{\emph{IEEE Transactions on Knowledge and Data Engineering}} \bibinfo{volume}{29}, \bibinfo{number}{1} (\bibinfo{year}{2017}), \bibinfo{pages}{17--37}.
\newblock


\bibitem[Shi et~al\mbox{.}(2020)]%
        {interpret1}
\bibfield{author}{\bibinfo{person}{Jinghan Shi}, \bibinfo{person}{Houye Ji}, \bibinfo{person}{Chuan Shi}, \bibinfo{person}{Xiao Wang}, \bibinfo{person}{Zhiqiang Zhang}, {and} \bibinfo{person}{Jun Zhou}.} \bibinfo{year}{2020}\natexlab{}.
\newblock \showarticletitle{Heterogeneous graph neural network for recommendation}. In \bibinfo{booktitle}{\emph{ICML Workshop}}.
\newblock


\bibitem[Song et~al\mbox{.}(2021)]%
        {social2}
\bibfield{author}{\bibinfo{person}{Changhao Song}, \bibinfo{person}{Bo Wang}, \bibinfo{person}{Qinxue Jiang}, \bibinfo{person}{Yehua Zhang}, \bibinfo{person}{Ruifang He}, {and} \bibinfo{person}{Yuexian Hou}.} \bibinfo{year}{2021}\natexlab{}.
\newblock \showarticletitle{Social Recommendation with Implicit Social Influence}. In \bibinfo{booktitle}{\emph{Proceedings of the 44th International ACM SIGIR Conference on Research and Development in Information Retrieval}}. \bibinfo{pages}{1788–1792}.
\newblock


\bibitem[Van~der Maaten and Hinton(2008)]%
        {tsne}
\bibfield{author}{\bibinfo{person}{Laurens Van~der Maaten} {and} \bibinfo{person}{Geoffrey Hinton}.} \bibinfo{year}{2008}\natexlab{}.
\newblock \showarticletitle{Visualizing data using t-SNE.}
\newblock \bibinfo{journal}{\emph{Journal of machine learning research}} \bibinfo{volume}{9}, \bibinfo{number}{11} (\bibinfo{year}{2008}).
\newblock


\bibitem[Wang et~al\mbox{.}(2019a)]%
        {ngcf}
\bibfield{author}{\bibinfo{person}{Xiang Wang}, \bibinfo{person}{Xiangnan He}, \bibinfo{person}{Meng Wang}, \bibinfo{person}{Fuli Feng}, {and} \bibinfo{person}{Tat-Seng Chua}.} \bibinfo{year}{2019}\natexlab{a}.
\newblock \showarticletitle{Neural Graph Collaborative Filtering}. In \bibinfo{booktitle}{\emph{Proceedings of the 42nd International ACM SIGIR Conference on Research and Development in Information Retrieval}}. \bibinfo{pages}{165–174}.
\newblock


\bibitem[Wang et~al\mbox{.}(2019b)]%
        {han}
\bibfield{author}{\bibinfo{person}{Xiao Wang}, \bibinfo{person}{Houye Ji}, \bibinfo{person}{Chuan Shi}, \bibinfo{person}{Bai Wang}, \bibinfo{person}{Yanfang Ye}, \bibinfo{person}{Peng Cui}, {and} \bibinfo{person}{Philip~S Yu}.} \bibinfo{year}{2019}\natexlab{b}.
\newblock \showarticletitle{Heterogeneous Graph Attention Network}. In \bibinfo{booktitle}{\emph{The World Wide Web Conference}}. \bibinfo{pages}{2022–2032}.
\newblock


\bibitem[Wang and Zhang(2022)]%
        {powerfulsgnn}
\bibfield{author}{\bibinfo{person}{Xiyuan Wang} {and} \bibinfo{person}{Muhan Zhang}.} \bibinfo{year}{2022}\natexlab{}.
\newblock \showarticletitle{How powerful are spectral graph neural networks}. In \bibinfo{booktitle}{\emph{International Conference on Machine Learning}}. PMLR, \bibinfo{pages}{23341--23362}.
\newblock


\bibitem[Wang et~al\mbox{.}(2020b)]%
        {amgcn}
\bibfield{author}{\bibinfo{person}{Xiao Wang}, \bibinfo{person}{Meiqi Zhu}, \bibinfo{person}{Deyu Bo}, \bibinfo{person}{Peng Cui}, \bibinfo{person}{Chuan Shi}, {and} \bibinfo{person}{Jian Pei}.} \bibinfo{year}{2020}\natexlab{b}.
\newblock \showarticletitle{Am-gcn: Adaptive multi-channel graph convolutional networks}. In \bibinfo{booktitle}{\emph{Proceedings of the 26th ACM SIGKDD International conference on knowledge discovery and data mining}}. \bibinfo{pages}{1243--1253}.
\newblock


\bibitem[Wang et~al\mbox{.}(2020a)]%
        {interpret2}
\bibfield{author}{\bibinfo{person}{Yifan Wang}, \bibinfo{person}{Suyao Tang}, \bibinfo{person}{Yuntong Lei}, \bibinfo{person}{Weiping Song}, \bibinfo{person}{Sheng Wang}, {and} \bibinfo{person}{Ming Zhang}.} \bibinfo{year}{2020}\natexlab{a}.
\newblock \showarticletitle{Disenhan: Disentangled heterogeneous graph attention network for recommendation}. In \bibinfo{booktitle}{\emph{Proceedings of the 29th ACM International Conference on Information and Knowledge Management}}.
\newblock


\bibitem[Welling and Kipf(2016)]%
        {gcn}
\bibfield{author}{\bibinfo{person}{Max Welling} {and} \bibinfo{person}{Thomas~N Kipf}.} \bibinfo{year}{2016}\natexlab{}.
\newblock \showarticletitle{Semi-supervised classification with graph convolutional networks}. In \bibinfo{booktitle}{\emph{International Conference on Learning Representations}}.
\newblock


\bibitem[Wu et~al\mbox{.}(2017)]%
        {rrn}
\bibfield{author}{\bibinfo{person}{Chao-Yuan Wu}, \bibinfo{person}{Amr Ahmed}, \bibinfo{person}{Alex Beutel}, \bibinfo{person}{Alexander~J Smola}, {and} \bibinfo{person}{How Jing}.} \bibinfo{year}{2017}\natexlab{}.
\newblock \showarticletitle{Recurrent recommender networks}. In \bibinfo{booktitle}{\emph{Proceedings of the tenth ACM international conference on web search and data mining}}.
\newblock


\bibitem[Wu et~al\mbox{.}(2020)]%
        {social1}
\bibfield{author}{\bibinfo{person}{Le Wu}, \bibinfo{person}{Junwei Li}, \bibinfo{person}{Peijie Sun}, \bibinfo{person}{Richang Hong}, \bibinfo{person}{Yong Ge}, {and} \bibinfo{person}{Meng Wang}.} \bibinfo{year}{2020}\natexlab{}.
\newblock \showarticletitle{Diffnet++: A neural influence and interest diffusion network for social recommendation}.
\newblock \bibinfo{journal}{\emph{IEEE Transactions on Knowledge and Data Engineering}} (\bibinfo{year}{2020}).
\newblock


\bibitem[Wu et~al\mbox{.}(2019)]%
        {wu2019neural}
\bibfield{author}{\bibinfo{person}{Le Wu}, \bibinfo{person}{Peijie Sun}, \bibinfo{person}{Yanjie Fu}, \bibinfo{person}{Richang Hong}, \bibinfo{person}{Xiting Wang}, {and} \bibinfo{person}{Meng Wang}.} \bibinfo{year}{2019}\natexlab{}.
\newblock \showarticletitle{A neural influence diffusion model for social recommendation}. In \bibinfo{booktitle}{\emph{Proceedings of the 42nd international ACM SIGIR conference on research and development in information retrieval}}. \bibinfo{pages}{235--244}.
\newblock


\bibitem[Wu et~al\mbox{.}(2022)]%
        {survey1}
\bibfield{author}{\bibinfo{person}{Shiwen Wu}, \bibinfo{person}{Fei Sun}, \bibinfo{person}{Wentao Zhang}, \bibinfo{person}{Xu Xie}, {and} \bibinfo{person}{Bin Cui}.} \bibinfo{year}{2022}\natexlab{}.
\newblock \showarticletitle{Graph Neural Networks in Recommender Systems: A Survey}.
\newblock \bibinfo{journal}{\emph{ACM Comput. Surv.}} (\bibinfo{year}{2022}).
\newblock


\bibitem[Xia et~al\mbox{.}(2022)]%
        {fire}
\bibfield{author}{\bibinfo{person}{Jiafeng Xia}, \bibinfo{person}{Dongsheng Li}, \bibinfo{person}{Hansu Gu}, \bibinfo{person}{Jiahao Liu}, \bibinfo{person}{Tun Lu}, {and} \bibinfo{person}{Ning Gu}.} \bibinfo{year}{2022}\natexlab{}.
\newblock \showarticletitle{FIRE: Fast Incremental Recommendation with Graph Signal Processing}. In \bibinfo{booktitle}{\emph{Proceedings of the ACM Web Conference 2022}}. \bibinfo{pages}{2360–2369}.
\newblock


\bibitem[Xia et~al\mbox{.}(2021)]%
        {igcn}
\bibfield{author}{\bibinfo{person}{Jiafeng Xia}, \bibinfo{person}{Dongsheng Li}, \bibinfo{person}{Hansu Gu}, \bibinfo{person}{Tun Lu}, \bibinfo{person}{Peng Zhang}, {and} \bibinfo{person}{Ning Gu}.} \bibinfo{year}{2021}\natexlab{}.
\newblock \showarticletitle{Incremental Graph Convolutional Network for Collaborative Filtering}. In \bibinfo{booktitle}{\emph{Proceedings of the 30th ACM International Conference on Information and Knowledge Management}}. \bibinfo{pages}{2170–2179}.
\newblock


\bibitem[Xia et~al\mbox{.}(2024a)]%
        {fagsp}
\bibfield{author}{\bibinfo{person}{Jiafeng Xia}, \bibinfo{person}{Dongsheng Li}, \bibinfo{person}{Hansu Gu}, \bibinfo{person}{Tun Lu}, \bibinfo{person}{Peng Zhang}, \bibinfo{person}{Li Shang}, {and} \bibinfo{person}{Ning Gu}.} \bibinfo{year}{2024}\natexlab{a}.
\newblock \showarticletitle{Frequency-aware Graph Signal Processing for Collaborative Filtering}.
\newblock \bibinfo{journal}{\emph{arXiv preprint arXiv:2402.08426}} (\bibinfo{year}{2024}).
\newblock


\bibitem[Xia et~al\mbox{.}(2024b)]%
        {higsp}
\bibfield{author}{\bibinfo{person}{Jiafeng Xia}, \bibinfo{person}{Dongsheng Li}, \bibinfo{person}{Hansu Gu}, \bibinfo{person}{Tun Lu}, \bibinfo{person}{Peng Zhang}, \bibinfo{person}{Li Shang}, {and} \bibinfo{person}{Ning Gu}.} \bibinfo{year}{2024}\natexlab{b}.
\newblock \showarticletitle{Hierarchical Graph Signal Processing for Collaborative Filtering}. In \bibinfo{booktitle}{\emph{Proceedings of the ACM on Web Conference 2024}}. \bibinfo{pages}{3229--3240}.
\newblock


\bibitem[Xia et~al\mbox{.}(2024c)]%
        {neufilter}
\bibfield{author}{\bibinfo{person}{Jiafeng Xia}, \bibinfo{person}{Dongsheng Li}, \bibinfo{person}{Hansu Gu}, \bibinfo{person}{Tun Lu}, \bibinfo{person}{Peng Zhang}, \bibinfo{person}{Li Shang}, {and} \bibinfo{person}{Ning Gu}.} \bibinfo{year}{2024}\natexlab{c}.
\newblock \showarticletitle{Neural Kalman Filtering for Robust Temporal Recommendation}. In \bibinfo{booktitle}{\emph{Proceedings of the 17th ACM International Conference on Web Search and Data Mining}}. \bibinfo{pages}{836--845}.
\newblock


\bibitem[Xie et~al\mbox{.}(2021)]%
        {interpret4}
\bibfield{author}{\bibinfo{person}{Qianqian Xie}, \bibinfo{person}{Yutao Zhu}, \bibinfo{person}{Jimin Huang}, \bibinfo{person}{Pan Du}, {and} \bibinfo{person}{Jian-Yun Nie}.} \bibinfo{year}{2021}\natexlab{}.
\newblock \showarticletitle{Graph neural collaborative topic model for citation recommendation}.
\newblock \bibinfo{journal}{\emph{ACM Transactions on Information Systems (TOIS)}} \bibinfo{volume}{40}, \bibinfo{number}{3} (\bibinfo{year}{2021}).
\newblock


\bibitem[Xin et~al\mbox{.}(2020)]%
        {oversmoothingvisualization}
\bibfield{author}{\bibinfo{person}{Xin Xin}, \bibinfo{person}{Alexandros Karatzoglou}, \bibinfo{person}{Ioannis Arapakis}, {and} \bibinfo{person}{Joemon~M Jose}.} \bibinfo{year}{2020}\natexlab{}.
\newblock \showarticletitle{Graph highway networks}.
\newblock \bibinfo{journal}{\emph{arXiv preprint arXiv:2004.04635}} (\bibinfo{year}{2020}).
\newblock


\bibitem[Xu et~al\mbox{.}(2018)]%
        {gin}
\bibfield{author}{\bibinfo{person}{Keyulu Xu}, \bibinfo{person}{Weihua Hu}, \bibinfo{person}{Jure Leskovec}, {and} \bibinfo{person}{Stefanie Jegelka}.} \bibinfo{year}{2018}\natexlab{}.
\newblock \showarticletitle{How powerful are graph neural networks?}. In \bibinfo{booktitle}{\emph{International Conference on Learning Representations}}.
\newblock


\bibitem[Yang et~al\mbox{.}(2021)]%
        {cpf}
\bibfield{author}{\bibinfo{person}{Cheng Yang}, \bibinfo{person}{Jiawei Liu}, {and} \bibinfo{person}{Chuan Shi}.} \bibinfo{year}{2021}\natexlab{}.
\newblock \showarticletitle{Extract the knowledge of graph neural networks and go beyond it: An effective knowledge distillation framework}. In \bibinfo{booktitle}{\emph{Proceedings of the Web Conference}}.
\newblock


\bibitem[Ye et~al\mbox{.}(2023)]%
        {gnncftref1}
\bibfield{author}{\bibinfo{person}{Haibo Ye}, \bibinfo{person}{Xinjie Li}, \bibinfo{person}{Yuan Yao}, {and} \bibinfo{person}{Hanghang Tong}.} \bibinfo{year}{2023}\natexlab{}.
\newblock \showarticletitle{Towards robust neural graph collaborative filtering via structure denoising and embedding perturbation}.
\newblock \bibinfo{journal}{\emph{ACM Transactions on Information Systems}} \bibinfo{volume}{41}, \bibinfo{number}{3} (\bibinfo{year}{2023}), \bibinfo{pages}{1--28}.
\newblock


\bibitem[Zhang et~al\mbox{.}(2020)]%
        {zhang2020explainable}
\bibfield{author}{\bibinfo{person}{Yongfeng Zhang}, \bibinfo{person}{Xu Chen}, {et~al\mbox{.}}} \bibinfo{year}{2020}\natexlab{}.
\newblock \showarticletitle{Explainable recommendation: A survey and new perspectives}.
\newblock \bibinfo{journal}{\emph{Foundations and Trends{\textregistered} in Information Retrieval}} \bibinfo{volume}{14}, \bibinfo{number}{1} (\bibinfo{year}{2020}), \bibinfo{pages}{1--101}.
\newblock


\bibitem[Zhou et~al\mbox{.}(2021)]%
        {saffnnet}
\bibfield{author}{\bibinfo{person}{Zhen Zhou}, \bibinfo{person}{Yan Zhou}, \bibinfo{person}{Dongli Wang}, \bibinfo{person}{Jinzhen Mu}, {and} \bibinfo{person}{Haibin Zhou}.} \bibinfo{year}{2021}\natexlab{}.
\newblock \showarticletitle{Self-attention feature fusion network for semantic segmentation}.
\newblock \bibinfo{journal}{\emph{Neurocomputing}}  \bibinfo{volume}{453} (\bibinfo{year}{2021}), \bibinfo{pages}{50--59}.
\newblock


\end{thebibliography}

\appendix

\end{document}